\documentclass[floatfix,aps,prd,showkeys,twocolumn,nofootinbib, longbibliography]{revtex4-2}

\usepackage{amsmath}
\usepackage{amssymb}
\usepackage{xcolor}
\usepackage{graphicx}
\usepackage{dcolumn}
\usepackage[loop]{animate}
\usepackage{comment}
\usepackage[normalem]{ulem}
\usepackage[bottom]{footmisc}
\usepackage{multirow}
\usepackage{afterpage}
\usepackage{hyperref}

\newcommand{\bs}{\boldsymbol}
\newcommand{\p}{\partial}
\newcommand{\pnabla}{{}^\perp\nabla}

\newcommand{\const}{\text{const}}

\newcommand{\sgn}{\text{sgn}}
\newcommand{\frg}{\mathfrak{g}}
\newcommand{\fru}{\mathfrak{u}}
\newcommand{\rmg}{\mathrm{g}}
\newcommand{\mnras}{\text{MNRAS}}

\begin{document}

\title{Instability windows of relativistic $r$-modes\\
in stably stratified neutron stars with hyperonic cores}

\date{\today}

\author{K. Y. Kraav}
\author{M. E. Gusakov}
\author{E. M. Kantor}

\affiliation{Ioffe Institute, Polytekhnicheskaya 26, St.-Petersburg 194021, Russia}


\begin{abstract}

$R$-modes are oscillations in rotating stars, primarily restored by the Coriolis force. Among stellar oscillations, they are the most susceptible to the Chandrasekhar-Friedman-Schutz (CFS) instability driven by gravitational wave emission, making them promising targets for current and future gravitational wave searches. The development of this instability, however, requires it to overcome dissipative processes within the star. As a result, $r$-modes become unstable only for certain combinations of stellar angular velocity $\Omega$ and (redshifted) temperature $T^\infty$, defining the so-called instability window on the $(\Omega, \, T^\infty)$ plane. At high temperatures, bulk viscosity $\zeta$, arising from out-of-equilibrium chemical reactions, is the dominant dissipative mechanism suppressing the CFS instability. Dissipation due to $\zeta$ can be greatly enhanced by two independent mechanisms: (1) the presence of hyperons, which significantly increases the bulk viscosity, and (2) the distinctive properties of relativistic $r$-modes in nonbarotropic matter, which further amplify dissipation beyond Newtonian predictions. In this work, we present the first investigation of the combined impact of these two mechanisms on $r$-mode instability windows. Our calculations also account for the fact that chemical reactions modify the adiabatic index, in addition to producing bulk viscosity. We further estimate the influence of nucleon superconductivity and superfluidity on the instability windows. By comparing our predictions with recent observations of neutron stars in low-mass X-ray binaries, we find that bulk viscosity in hyperonic matter may provide the necessary dissipation to stabilize $r$-modes in the fastest-spinning and moderately hot stars, even when nucleon superfluidity and superconductivity are taken into account. These results have important implications for the interpretation of observations and for the broader understanding of relativistic $r$-mode physics.
\end{abstract}

\maketitle


\section{Introduction}

Neutron stars (NS) are often described as natural laboratories for investigating matter under extreme conditions, unattainable in terrestrial experiments. Typically, they have masses $M$ of about $1.4$ solar masses $M_\odot$ and are confined within a radius $R$ of approximately $12 \, \rm km$, indicating that the densities in their interiors may exceed those found in atomic nuclei. At the same time, the interior temperatures $T$ of not too hot/young neutron stars usually fall below $10^9\, \rm K$, which, owing to the high densities, is significantly lower than the Fermi temperatures $T_{\rm F}\sim 10^{12}-10^{13} \, \rm K$ of the matter constituents (neutrons, protons, electrons, muons, and, possibly, other particle species). The lack of experimental data, combined with the absence of a precise theory that accounts for strong nuclear many-body interactions, has led to the development of a handful of theoretical models, each attempting to predict the properties of such ``cold" ultradense matter. Hopefully, some of these models can be favored over others by comparing their predictions with observations of neutron stars. In particular, it is believed that future gravitational-wave (GW) and electromagnetic observations of the NS oscillations can be employed to clarify the properties of the matter in their interiors.

A perturbed NS can exhibit a variety of oscillation modes, which are conventionally classified according to the dominant restoring force that governs their dynamics \cite{cowling1941}. The $r$-modes (also known as Rossby waves), arising in rotating stars and primarily restored by the Coriolis force, are of particular interest to gravitational-wave astronomy (see, e.g., \cite{zhuetal2019, cppo2022, hhcc2019, roci2021, abbotetal2021, boztepeetal2020, bhattacharyaetal2017, carideetal2019} and references therein). Of all oscillation modes, they appear to be the most susceptible to Chandrasekhar-Friedman-Schutz (CFS) instability with respect to gravitational-wave emission \cite{chandra1970, fs1978_1, fs1978_2, friedman1978}. In the case of $r$-modes, CFS mechanism suggests that, at an {\it arbitrary} NS rotation rate, the GW radiation, produced during oscillations, causes the transfer of the stellar rotation energy into the oscillation energy  \cite{andersson1998, fm1998, chugunov2017}. Under favorable conditions, this process can lead to the growth of the mode amplitude, thereby enhancing GW emission, which completes the positive feedback loop.

In reality, various dissipative mechanisms (e.g., viscosity, particle diffusion, and others) cause the conversion of the $r$-mode energy into heat, which, in contrast to CFS instability, tends to dampen the oscillation amplitude, as reviewed in \cite{haskell14, glampedakis2018}. As a result, the development of the instability depends on whether the CFS mechanism is sufficiently strong to overcome the oscillation energy losses caused by dissipation. Usually, the power $\dot{E}_{\rm GW}(\Omega)$ of CFS mechanism depends only on the NS rotation rate $\Omega$, while the dissipative energy losses $\dot{E}_{\rm diss}(\Omega,T^\infty)$ depend additionally on the (redshifted) stellar temperature $T^\infty$. By comparing $\dot{E}_{\rm GW}$ against $\dot{E}_{\rm diss}$, one may determine the so-called {\it $r$-mode instability window} -- the region on the $(\Omega,T^\infty)$ plane, where the CFS mechanism prevails over the dissipative energy drain, and the $r$-mode becomes unstable \cite{low1998, levin1999, bildstenetal2000, ak2001, cgk2017}. 

\begin{figure}
\centering
\includegraphics[width=1.0\linewidth]{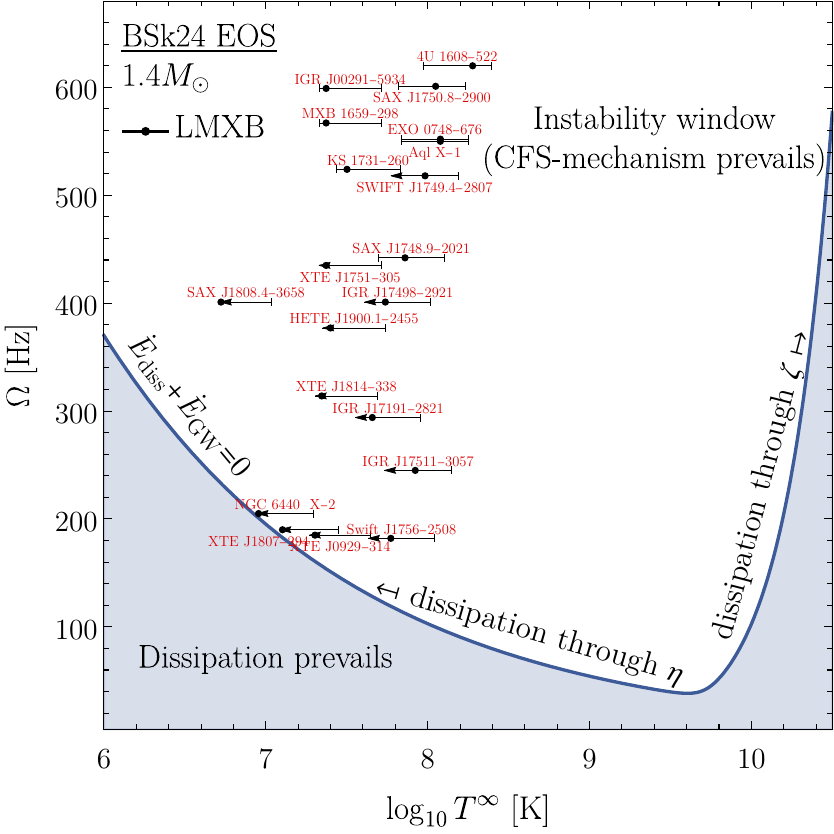}
\caption{Typical instability window of Newtonian $r$-modes in nonbarotropic neutron stars with nucleonic composition. Here, shear $(\eta)$ and bulk $(\zeta)$ viscosities are the only sources of dissipation. The blue curve divides the plane into two regions: the lower one, where $r$-modes are suppressed by dissipation, and the upper one -- the instability window -- where the CFS mechanism prevails and $r$-modes become unstable. Black dots display LMXB observational data from \cite{kgd2021}, while the error bars illustrate the uncertainty in temperature arising from the poorly known composition of the outer NS layers (see \cite{gck2014} for details). The figure is drawn using our previous results reported in \cite{kgk2024}.}
\label{representative window}
\end{figure}

The shape of the {\it modeled} instability window depends on multiple factors, including the assumed equation of state (EOS) of the matter, its chemical composition, the dissipative mechanisms considered, and the techniques and approximations adopted in the calculations. For the reasons to be explained shortly, almost all calculations of the instability windows employ $r$-modes, modeled within the framework of Newtonian gravity rather than general relativity (GR). The simplest, or, as we shall say, {\it minimal models}, following the pioneering study \cite{low1998}, assume a nucleonic composition of the NS core (i.e., not including hyperons or other exotic particles) and describe the matter using a {\it barotropic} EOS, which treats pressure and other thermodynamic quantities as functions of a single variable (e.g., energy density or baryon number density). As for dissipative mechanisms, such models typically account only for {\it shear viscosity} $(\eta)$ [i.e., friction due to relative motion of fluid layers], and {\it bulk viscosity} $(\zeta)$ [i.e., heat production due to deviations from chemical equilibrium caused by fluid compression/decompression]. The former stabilizes $r$-modes at low temperatures, while the latter is efficient at high temperatures $T^\infty>10^9 \, \rm K$, which results into the V-shaped boundary of the instability window. 

Figure~\ref{representative window} illustrates a representative instability window of Newtonian $r$-modes in {\it nonbarotropic} neutron stars with nucleonic composition (borrowed from our preceding study~\cite{kgk2024}). Although the calculation assumes the nonbarotropic matter, the shape of the window is qualitatively the same as in the previously described minimal models, treating the matter as barotropic. The blue curve divides the $(\Omega,T^\infty)$ plane into two regions: the lower shaded region, where $r$-modes are suppressed by dissipation, and the instability window located above the curve. Black dots display the observational data of neutron stars in low-mass X-ray binaries (LMXBs) taken from \cite{kgd2021}, while the error bars show the uncertainty in temperature arising from  poorly known composition of the outer NS layers (see \cite{gck2014} for details). Theoretically, finding an NS with excited $r$-mode inside the instability window should be highly improbable, since, due to rapid spin-down by gravitational-wave emission, the star spends therein only a tiny fraction of its lifetime (see, e.g., \cite{owenetal1998, levin1999, hl2000, gck2014, haskell14}). Therefore, the fact that numerous stars in LMXBs reside within the described instability windows necessitates a modification of the theory. The most popular approach is to look for additional, previously unaccounted-for channels of energy dissipation that would be strong enough to reshape the instability window into a new form that does not include the observed sources. Dissipation in the Ekman layer \cite{bu2000, lu2001, ga2006}, mutual friction \cite{hap2009, lm2000, ly2003, pha2009, haskell14}, resonant $r$-mode stabilization by superfluid modes \cite{gck14a, gck2014, kg2017, dkg2019, kgd2020, kgd2021}, particle diffusion \cite{kgk2021, kgk2024}, and other mechanisms have been considered in the literature as possible candidates (see reviews \cite{glampedakis2018, haskell14} for details). An alternative option is to improve the modeling of $r$-modes by, for example, using more realistic EOSs or including relativistic effects.

In this study, we consider two modifications to the minimal instability window model. First, {\it we allow a neutron star to contain hyperons in its core}. In stars with a nucleonic chemical composition, the primary contribution to bulk viscosity arises from modified Urca processes and, in the most massive stars, from direct Urca processes \cite{haenseletal2000, haenseletal2001}. At low temperatures, $T\ll T_{\rm F}$, the rates of these reactions are strongly suppressed as $(T/T_{\rm F})^6$ and $(T/T_{\rm F})^4$, respectively (assuming nonsuperfluid matter). As a result, bulk viscosity in {\it nucleonic} stars can substantially contribute to the $r$-mode energy dissipation only at relatively high temperatures (see Fig.\,\ref{representative window}). In stars with a {\it hyperonic} composition, however, bulk viscosity is primarily determined by weak nonleptonic processes involving hyperons (see \cite{ofengeimetal2019} and references therein). At low temperatures, the rates of these reactions are suppressed only as $(T/T_{\rm F})^2$, implying that they proceed substantially faster than Urca processes and, therefore, lead to larger values of bulk viscosity%
\footnote{
Strictly speaking, faster reactions correspond to larger bulk viscosity only when relaxation to chemical equilibrium takes significantly longer than the oscillation period. In the opposite limit of fast relaxation, higher reaction rates imply smaller bulk viscosity (see Sec.~\ref{Bulk viscosity and adiabatic index}).
}.
Consequently, the energy dissipation due to bulk viscosity becomes strongly amplified by the presence of hyperons. Until recently, the $r$-mode instability window calculations have predicted this effect to be insufficiently strong to stabilize the $r$-modes in the observed NSs \cite{lo2002, nb2006, rb2003, haskell14, vidanya2015}. However, these calculations assumed a $\Sigma^-\Lambda$ hyperonic composition of the matter, while various modern equations of state (e.g., \cite{ghk14, rsw2018, negreirosetal2018, fortinetal2017, providenciaetal2019}) suggest that $\Lambda$- and $\Xi^-$-hyperons are likely the first to emerge with increasing density. Moreover, these calculations underestimated the reaction rates of weak nonleptonic processes by neglecting their most efficient operating channel, namely, through meson exchange \cite{dd04}. As was demonstrated in \cite{ofengeimetal2019}, accounting for this channel, along with replacing the $\Sigma^-\Lambda$ hyperonic composition with the $\Lambda\Xi^-$ one, makes hyperonic bulk viscosity a substantially more viable solution to the LMXB paradox than previously thought. Specifically, for Newtonian $r$-modes, it was shown that, for certain combinations of equations of state and stellar masses, hyperonic bulk viscosity allows to stabilize the observed sources. However, as we recently discovered \cite{kgk2022_1, kgk2022_2, kgk2024}, the Newtonian framework appears inadequate for describing $r$-mode properties under physical conditions typical of NS interiors.

It turns out that {\it realistic $r$-mode modeling requires simultaneously accounting for the effects of GR and treating the stellar matter as nonbarotropic, which is the second modification we suggest to consider}. Surprisingly, relativistic $r$-modes in nonbarotropic neutron stars demonstrate rather peculiar properties, compared to their Newtonian cousins \cite{kgk2022_1, kgk2022_2}. In GR, a massive star, rotating with angular velocity $\Omega$, entrains the local inertial reference frames at a radial coordinate $r$ with angular velocity $\Omega\omega(r)$, as measured by a distant inertial observer \cite{hartle1967}%
\footnote{
The factor $\Omega$ is usually absorbed in the definition of $\omega(r)$, and one writes $\omega(r)$ instead of $\Omega\omega(r)$. In this study, for convenience, we deliberately factor $\Omega$ out, ensuring that $\omega(r)$ is dimensionless and does not depend on $\Omega$.
}.
In nonbarotropic stars, this phenomenon, known as the frame-dragging effect, modifies $r$-mode oscillation spectrum, influences the geometry of the matter flows and leads to nonanalytic%
\footnote{
In this paper, we use the terms ``nonanalytic" and ``nonanalyticity" to indicate that the discussed object (typically, $r$-mode eigenfunctions) is described by {\it nonanalytic functions of $\Omega$} (and, therefore, cannot be Taylor-expanded in $\Omega$).
}
dependence of their eigenfunctions (flow velocity, pressure perturbation, energy density perturbation, etc.) on the NS angular velocity $\Omega$. This is very different from the analytic behavior of $r$-modes in Newtonian theory and leads, apart from other manifestations, to the exponential suppression of relativistic $r$-modes in the NS core as $\Omega$ approaches zero. At least in nucleonic matter, these peculiarities were shown to substantially reshape the $r$-mode instability windows and, surprisingly, result in an amplification of the $r$-mode damping due to bulk viscosity by at least two orders of magnitude \cite{kgk2024}.

Summarizing, the dissipation due to bulk viscosity can be substantially enhanced by two independent physical mechanisms: the amplification resulting from the peculiar properties of relativistic $r$-modes and the modification of the bulk viscosity due to the presence of hyperons, if allowed by the EOS. In this study, we examine for the first time the combined effect of these two factors on the $r$-mode instability windows. Essentially, we apply the previously developed theory of nonanalytic relativistic $r$-modes \cite{kgk2022_1, kgk2022_2} and their instability windows \cite{kgk2024} to a particular case of neutron stars with hyperonic cores, where bulk viscosity is provided mainly by weak nonleptonic processes, as described in \cite{ofengeimetal2019}. To make the narrative reasonably self-contained, we begin the paper with a theoretical overview (Sec.~\ref{Theoretical overview}) that summarizes the key general ideas and approaches used in the calculations. The overview touches upon the adopted NS model \eqref{Neutron star hydrodynamical model}, properties of NS equilibrium \eqref{Neutron star equilibrium}, hydrodynamic modeling of oscillations in the presence of chemical reactions \eqref{NS oscillations}, methods for estimating the effects of nucleon superfluidity and superconductivity \eqref{SfSc effects}, and, finally, the concept of instability windows \eqref{Instability windows}. After the overview, we turn to application of the theory to relativistic $r$-modes in neutron stars with hyperonic composition (Sec.~\ref{Application to r-modes}). We discuss the relativistic $r$-mode peculiarities \eqref{Nonanalytic r-modes} and their influence on the $r$-mode amplification by CFS mechanism and viscous damping \eqref{Energy change rates}. Then we provide the results of our numerical calculations of instability windows and compare them against the LMXB observational data \eqref{Numerical results}. The final section \eqref{Conclusion} contains some concluding remarks. 

Throughout the text, we use standard notations for the speed of light $c$, the gravitational constant $G$, and the Boltzmann constant $k_{\rm B}$. We use Latin indices $(k,m,n,\dots)$ to label physical quantities $f_k$, associated with particle species $k$, while Greek indices $(\mu,\nu,\rho,\dots)$ label tensor components. For convenience, we enclose Latin indices in brackets for quantities $f_{(k)}^\mu$ that contain both Latin and Greek indices. We also reserve notations $\frg_{\mu\nu}$ and $\nabla_\mu$ for the metric tensor and the associated covariant derivative, respectively. Finally, we adopt the $\{-,+,+,+\}$ convention for the metric tensor signature and, unless stated otherwise, imply summation over all repeated indices, both Greek and Latin. The four-velocity $\fru^\mu$ is normalized to $-1$ by the condition $\fru^\mu\fru_\mu=-1$. 


\section{Theoretical overview}\label{Theoretical overview}

This section offers a brief overview of the theoretical framework used to investigate relativistic $r$-modes and their instability windows in hyperonic stars. Our approach is primarily based on the recently developed relativistic dissipative hydrodynamics of normal \cite{dommes2020} and superfluid/superconducting mixtures \cite{dg2021}, along with its applications \cite{kgk2022_1, kgk2022_2, kgk2024} to modeling relativistic $r$-modes. To describe the influence of chemical reactions on NS oscillations, we closely follow the approach of \cite{ofengeimetal2019} and extend their considerations by incorporating relativistic effects and discussing how chemical reactions modify the adiabatic index of matter. To estimate the effects of proton superconductivity and neutron superfluidity, we employ the formalism of reduction factors, as described, e.g., in \cite{haenseletal2000, haenseletal2001, haenseletal2002}.


\subsection{Neutron star hydrodynamical model}\label{Neutron star hydrodynamical model}

In this study, we employ a simplified NS model, presented in \cite{kgk2022_1}. Specifically, we divide the stellar interior into two physically distinct regions: the outer region, corresponding to the NS crust, and the inner one, corresponding to the NS core. We neglect the stellar magnetic field, as its effect on the hydrodynamical large-scale NS oscillations, including $r$-modes \cite{chirenti2013, kinney2003}, is likely negligible under conditions typical for LMXBs. Throughout the star, the matter is assumed to be degenerate and quasineutral, meaning it maintains approximately zero electrical charge and current.

Typically, the crust is quite thin ($\sim 0.1 R$), and almost the entire mass of the star is concentrated in its core. For this reason, the crust can be treated approximately while still providing an adequate description of {\it global} NS oscillations, spanning both core and crust. Here, we neglect the elastic properties of the crust, which allows us to treat the entire star hydrodynamically, i.e., as a liquid mixture of various particle species. This approximation is valid for studying high-frequency hydrodynamic oscillations, when the contributions of the crustal lattice to the oscillation equations are relatively small or, equivalently, when oscillation frequency exceeds the frequencies of crustal torsional modes. This condition is partially met by the $r$-modes in LMXBs, as their frequencies likely exceed the fundamental and low-lying overtones of the crustal modes. The interaction of $r$-modes with high-overtone crustal modes, strictly speaking, may take place, as their frequencies may reach values on the order of hundreds of hertz, relevant for $r$-modes in LMXBs. Here, for simplicity, we ignore this effect, and, for a discussion on the influence of the crust-core interaction on $r$-mode properties,  refer the interested reader to  \cite{bu2000, yl2001, lu2001, ga2006}. In this study, we also treat the crust as {\it barotropic}, which means that the pressure $p$ (and other thermodynamic functions) can be considered as a function depending solely on the energy density $\varepsilon$: $p=p(\varepsilon)$.

Unlike the crust, we treat the core as {\it nonbarotropic}, meaning that the pressure and other thermodynamic quantities cannot be considered as functions of a single variable. Specifically, we describe the core assuming an EOS of the form $\varepsilon=\varepsilon(\{n_k\})$, where $\{n_k\}$ represents the set of number densities $n_k$ corresponding to various particle species. The specific chemical composition of the core is determined by EOS under consideration. Here, we focus on EOSs that, in addition to neutrons $\rm(n)$, protons $\rm(p)$, electrons $\rm(e)$, and muons $\rm(\mu)$ -- the standard nucleonic composition -- also permit, at sufficiently high densities, the presence of hyperons: particles composed of up $\rm(u)$, down $\rm(d)$, and necessarily strange $\rm(s)$ quarks. Depending on the EOS, different hyperons may emerge in varying orders as density increases. Various modern equations of state \cite{ghk14, rsw2018, negreirosetal2018, fortinetal2017, providenciaetal2019} suggest that $\Lambda$-hyperons, followed by $\Xi^-$-hyperons, are likely the first to appear. In this study, we restrict our analysis to this particular order of hyperon appearance. We characterize each of the mentioned particle species $k$ by its electric charge $e_k$, baryon number $b_k$, and strangeness $s_k$, as summarized in Table \ref{particles}. It is also convenient to introduce the baryon number density $n_b$, strangeness number density $n_{\rm s}$, particle fractions $y_k$, and strangeness fraction $y_{\rm s}$:
\begin{gather}
\label{new variables}
n_b\equiv b_k n_k, \quad n_{\rm s}\equiv s_k n_k, \quad y_k \equiv \frac{n_k}{n_b}, \quad y_{\rm s} \equiv \frac{n_{\rm s}}{n_b}.
\end{gather}
%

\begin{table}[t]
    \centering
    \begin{tabular}{|c|c|c|c|c|c|}
    \hline
    Particle   & Type            & Composition & $e_k/e$ & $b_k$   & $s_k$   \\
    \hline\hline
    n          & baryon, nucleon &  udd        &  0    &   1     &   0      \\
    \hline
    p          & baryon, nucleon &  uud        &  1    &   1     &   0      \\
    \hline
    $\Lambda$  & baryon, hyperon &  uds        &  0    &   1     &  -1      \\ 
    \hline
    $\Xi^-$    & baryon, hyperon &  dss        & -1    &   1     &  -2     \\
    \hline
    e          & lepton          &   -         & -1    &   0     &   0      \\
    \hline
    $\mu$      & lepton          &   -         & -1    &   0     &   0     \\
    \hline
    \end{tabular}
    \vspace{-0.2cm}
    \caption{Properties of particles in NS core: type, quark composition, electric charge $e_k$ in units of elementary (positive) charge $e$, baryon number $b_k$, and strangeness $s_k$.}
    \label{particles}
\end{table}

\begin{figure}
\centering
\includegraphics[width=1.0\linewidth]{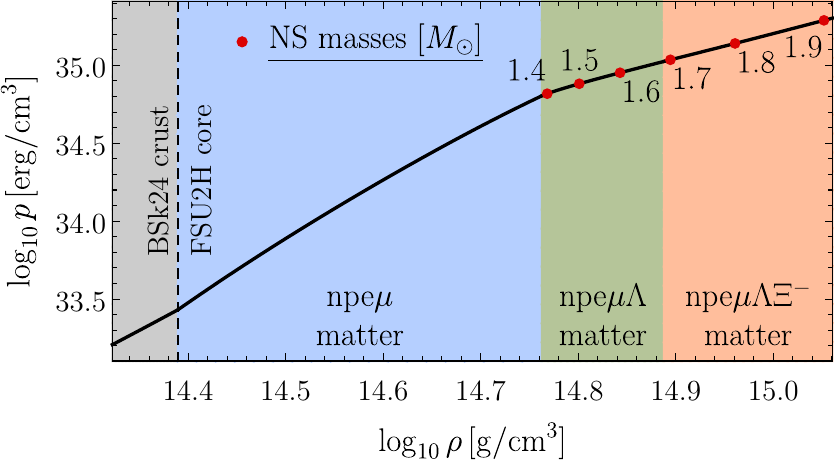}
\vspace{-0.6cm}
\caption{
The black curve shows the equilibrium pressure $p$ versus density $\rho$ 
for one of  the two considered equations of state. Red dots show central densities and pressures, corresponding to NS masses ranging from $1.4 M_\odot$ to $1.9 M_\odot$. Here, the crust is modeled using BSk24 EOS, while the core is described by FSU2H EOS (see details in Sec.\,\ref{Numerical results}).}
\label{nsmodel}
\vspace{-0.5cm}
\end{figure}

In our calculations, we model NS crust using BSk24 EOS, while for the core we employ either FSU2H or TM1C EOS (see details in Sec.\,\ref{Numerical results}). We determine the crust-core interface by the value $\rho_{\rm cc}$ of the density $\rho\equiv\varepsilon/c^2$ where the equilibrium pressure of the core EOS matches that of the crust EOS. Figure \,\ref{nsmodel} illustrates the resulting FSU2H core structure. The black curve shows the equilibrium pressure as a function of the density, while different colors show layers of the core with different chemical composition, ranging from nucleonic $\rm npe\mu$-matter to hyperonic $\rm npe\mu\Lambda\Xi^-$-matter. Red dots show central densities and pressures, corresponding to NS masses ranging from $1.4 \, M_\odot$ to $1.9 \, M_\odot$.

At temperatures below $10^9 \, \rm K$, typical for stars in LMXBs, neutrons and protons in NS interiors are likely to be in superfluid or superconducting states \cite{yls99, ls01, plps13}. Here, we allow protons to be either normal or strongly superconducting (i.e., when all protons form Cooper pairs). Similarly, neutrons will be treated either as normal or superfluid. Unlike proton superconductivity, however, we allow neutron superfluidity to be arbitrarily weak or strong.

Concerning hyperons, the current understanding of their superfluidity and superconductivity is considerably less developed. Some studies (e.g., \cite{ws2010, takatsukaetal2006}) based on \cite{takahashietal2001} tentatively indicate that $\Lambda$-hyperons may become superfluid only at sufficiently low temperatures, implying that under LMXB conditions they can be treated as normal. An alternative possibility that the critical temperature of $\Lambda$-hyperons may not be so low is also discussed in the literature and, of course, cannot be excluded with certainty (see the review \cite{sc2019} and references therein). Regarding $\Xi^-$-hyperons, there are indications that their superconductivity likely emerges at temperatures higher than those required for $\Lambda$-superfluidity, and it is quite possible that they exist in a superconducting state under NS conditions \cite{sc2019}. Nevertheless, in this study, for simplicity, we treat both $\Lambda$- and $\Xi^-$-hyperons as nonsuperfluid and nonsuperconducting. The effect of $\Xi^-$-superconductivity on the $r$-mode instability windows is expected to be weak, whereas $\Lambda$-superfluidity, in contrast, may have a drastic impact (see discussion in Sec.\,\ref{Conclusion}).

In our simplified model, we ignore the large-scale effects of nucleon superfluidity/superconductivity and describe the hydrodynamic motion of various particle species with a single collective four velocity $\fru^\mu$, as if the matter were normal. Thus, here the role of these effects is limited to modifying the kinetic properties of the matter, as will be reviewed in Sec.\,\ref{SfSc effects}. Notably, this approximation is fully justified not only for the matter with normal neutrons and protons, but also when neutrons are normal and protons are superconducting \cite{kgk2024}. This approximation, however, is less justified when neutrons are superfluid. Nevertheless, we still expect it to provide reasonable description of normal (nonsuperfluid) oscillation modes, where normal and superconducting components approximately comove.%
\footnote{
Note, however, that at certain rotation rates normal $r$-modes may experience avoided crossings with superfluid ones, and the resulting resonant interaction with the superfluid modes is expected to stabilize normal $r$-modes (see \cite{kgd2021} and references therein). Such resonant $r$-mode stabilization cannot be accounted for by our approach.
}

Now, we are in position to write down the general equations that govern NS dynamics. To this aim, we introduce the orthogonal projector $\perp^{\mu\nu}$ and orthogonal covariant derivative $\pnabla^\mu$, defined as follows:
\begin{gather}
\perp^{\mu\nu}\equiv\frg^{\mu\nu}+\fru^\mu\fru^\nu, \qquad \pnabla^\mu\equiv\perp^{\mu\nu}\nabla_\nu.
\end{gather}
Then, within approximations made, we can write the stress-energy tensor $T^{\mu\nu}$ and number density currents $j_{(k)}^\mu$ of various particle species as
\begin{gather}
\label{j and T}
j_{(k)}^\mu=n_k\fru^\mu, \quad T^{\mu\nu}=w \fru^\mu \fru^\nu + p \frg^{\mu\nu}-\eta c \sigma^{\mu\nu},
\end{gather}
where $w=p+\varepsilon$ is the enthalpy density, $\eta$ is the {\it shear viscosity} coefficient, and $\sigma^{\mu\nu}$ is the shear tensor:
\begin{gather}
\sigma^{\mu\nu}\equiv\pnabla^\mu \fru^\nu + \pnabla^\nu \fru^\mu-\frac{2}{3}\perp^{\mu\nu}\pnabla_\lambda \fru^\lambda.
\end{gather}
Using these expressions for $T^{\mu\nu}$ and $j_{(k)}^\mu$, it is straightforward to show that NS dynamics is described by the following equations:
\begin{gather}
\label{EulerEq}
w \fru^\mu\nabla_\mu \fru^\rho=-\pnabla^\rho p+\perp^\rho_\lambda \nabla_\mu(\eta c\sigma^{\mu\lambda}), \\
\label{ContEq}
\nabla_\mu(n_k\fru^\mu)=\Gamma_k/c, \\
\label{EOS}
\varepsilon=\varepsilon(\{n_k\}), \\
\label{thermodynamics}
d\varepsilon=\mu_k dn_k, \quad dp=n_k d\mu_k, \quad w\equiv p+\varepsilon=\mu_k n_k.
\end{gather}
where we introduced relativistic chemical potentials $\mu_k$ of various particle species. The first line \eqref{EulerEq} is the Euler equation $\perp^\rho_\lambda \nabla_\mu T^{\mu\lambda}=0$, the second \eqref{ContEq} is the set of continuity equations with sources $\Gamma_k$, powered by chemical reactions (to be discussed in detail in Secs.\ \ref{Neutron star equilibrium} and \ref{NS oscillations}), and remaining two lines \eqref{EOS}-\eqref{thermodynamics} represent the EOS and thermodynamic relations in degenerate matter.

Our equations support two physically distinct dissipative mechanisms. {\it First,} they account for the friction between fluid layers in relative motion. This effect is described by the last term in the stress energy tensor $T^{\mu\nu}$ \eqref{j and T} with underlying microphysics of the friction encoded in shear viscosity coefficient $\eta$. {\it Second,} our equations account for the energy losses due to out-of-equilibrium chemical reactions caused by fluid compression and rarefaction. The microphysics of this effect, in turn, is encoded in the sources $\Gamma_k$ that we explicitly include in continuity equations \eqref{ContEq}. When chemical reactions are sufficiently slow and almost do not affect chemical composition of the matter, this effect can be equivalently accounted for by discarding sources $\Gamma_k$ in continuity equations and adding to the stress-energy tensor $T^{\mu\nu}$ one more term
\begin{gather}
T^{\mu\nu}\to T^{\mu\nu}-\zeta c\perp^{\mu\nu}\pnabla_\lambda\fru^\lambda,
\end{gather}
where coefficient $\zeta$ is known as the {\it bulk viscosity coefficient}. However, when chemical reactions cannot be considered slow, introducing bulk viscosity does not fully account for their effect. As will be shown in Sec.\,\ref{NS oscillations}, apart from dissipative terms, responsible for bulk viscosity, sources $\Gamma_k$, generally, also produce nondissipative contributions to equations that have to be accounted for when reactions are sufficiently fast (to be more specific, chemical reactions modify the adiabatic index of the matter, see \cite{ghk14}). In the subsequent discussion,
up to a certain point, we follow the description in terms of sources $\Gamma_k$, since (within the approximations made) it provides the most accurate and transparent treatment of NS hydrodynamics in the presence of chemical reactions. When it comes to practical calculations of the energy dissipation caused by chemical reactions, we switch to more convenient description in terms of bulk viscosity (and modified adiabatic index). We will provide further details on how chemical reactions affect NS oscillations and, in particular, the relation between bulk viscosity, the adiabatic index, and the sources $\Gamma_k$ in Sec.~\ref{NS oscillations}. For now, we turn to a discussion of NS equilibrium, which will serve as the background for these oscillations.


\subsection{Neutron star equilibrium}\label{Neutron star equilibrium}

We define an equilibrium NS as being simultaneously in {\it hydrostatic}, {\it thermal}, and {\it chemical equilibrium}. In what follows, we label the corresponding equilibrium values $f_0$ of physical quantities $f$ by the index ``0".

{\it Chemical equilibrium} allows the use of the equilibrium EOS $p_0 = p_0(\varepsilon_0)$ in barotropic form throughout the entire equilibrium star, despite the nonbarotropic nature of its core. Indeed, in chemical equilibrium, the number densities $n_{k0}=n_{k0}(n_{b0})$ of various particle species in the core can be considered as depending solely on the baryon number density $n_{b0}$. This allows one to treat the pressure $p_0 = p_0({n_{k0}})$ and energy density $\varepsilon_0 = \varepsilon_0({n_{k0}})$ as functions of a single variable, $n_{b0}$, and therefore to treat the pressure in the equilibrium nonbarotropic core as a function of the energy density. To find number densities $n_{k0}$ at a given point of the equilibrium NS core, we follow \cite{hpy2007} and minimize the energy density $\varepsilon(\{n_k\})$, while ensuring the charge neutrality  $e_k n_k=0$ of the matter and maintaining the specified baryon number density $b_k n_k=n_b$, corresponding to the chosen point. Mathematically, this is achieved by introducing Lagrangian multipliers $a_{1,2}$ and minimizing the functional
\begin{gather}
\varepsilon(\{n_k\})-a_1 e_k n_k-a_2 (b_k n_k -n_b) \mapsto \rm min,
\end{gather}
which, given the EOS \eqref{EOS} and thermodynamic relations \eqref{thermodynamics}, leads (after simple algebraic manipulations) to
\begin{gather}
\label{composition}
\begin{gathered}
\mu_{{\rm n}0}b_k=\mu_{k0}+(e_k/e)\mu_{{\rm e}0}, \quad \mu_{k0}=\mu_{k0}(\{n_{m0}\}), \\
e_k n_{k0}=0, \quad b_k n_{k0}=n_{b0},
\end{gathered}
\end{gather}
These equations form a closed nonlinear system, which, as previously announced, allows finding the equilibrium number densities $n_{k0}=n_{k0}(n_{b0})$ as functions of baryon number density $n_{b0}$.

Under favorable physical conditions, particles in the NS core can participate in various chemical reactions. In nucleonic matter, the relevant reactions are the {\it weak leptonic processes}, mediated by weak interaction. These reactions include modified Urca (mUrca) processes \cite{haenseletal2001},
\begin{gather}
\label{wlpFirst}
n + N \rightarrow p + N + l + \bar{\nu}_l, \\
n + N + \nu_l \leftarrow p + N + l, 
\end{gather}
and, at sufficiently high densities, direct Urca (dUrca) processes \cite{haenseletal2000},
\begin{gather}
\label{wlpLast}
n\rightarrow p + l + \bar{\nu}_{\rm l}, \qquad n + \nu_l \leftarrow p+l, 
\end{gather}
where $N$ represents any nucleon, $l$ stands for any lepton, and $\nu_l/\bar{\nu}_l$ represent the corresponding neutrinos/antineutrinos, respectively. If hyperons are present, they can also participate in weak leptonic reactions analogous to the mUrca and dUrca processes. Moreover, in the case of hyperonic matter, another group of chemical reactions becomes available -- the so-called {\it weak nonleptonic processes}, mediated by weak or combined weak and strong interactions \cite{lo2002, haenseletal2002, dd04, ofengeimetal2019},
\begin{gather}
\label{wnpFirst}
n + n \leftrightarrow \Lambda + n, \\
\label{npLp}
n + p \leftrightarrow \Lambda + p, \\
\label{nLLL}
n + \Lambda \leftrightarrow \Lambda + \Lambda,  \\
\label{nXLX}
n + \Xi^- \leftrightarrow \Lambda + \Xi^-,  \\
\label{wnpLast}
n + \Lambda \leftrightarrow \Xi^- + p.
\end{gather}
Finally, under favorable conditions, hyperons can also participate in the (single available for $\rm npe\mu\Lambda\Xi^-$-matter) {\it strong process}, mediated by strong interaction,
\begin{gather}
\label{sp}
\Lambda + \Lambda  \leftrightarrow \Xi^- + p.
\end{gather}
It is worth keeping in mind that the listed chemical reactions occur only when permitted by the conservation laws of particle energy and momentum. Generally, for a given reaction, this may happen only above a certain threshold density, since the momenta of the involved particles are localized near their respective  Fermi surfaces.

For convenience, we have grouped all chemical reactions into pairs of direct $(\rightarrow)$ and inverse $(\leftarrow)$ processes. In each pair, the preferred ``direction" can be determined by the corresponding chemical imbalance, introduced as the difference in the chemical potentials of the ingoing (in) and the outgoing (out) particles involved in the direct process,
\begin{gather}
\label{chemical equilibrium}
\mu_{\rm in \leftrightarrow out}=\sum_{\rm in}\mu_k-\sum_{\rm out}\mu_k.
\end{gather}
The positive value of the chemical imbalance indicates that the direct process is preferred over the inverse one, and that the rate (i.e., the number of reactions per unit time per unit volume) of the direct reactions exceeds that of the inverse reactions. The negative value of the chemical imbalance, in turn, indicates the opposite case of the inverse process being preferred over the direct one. In chemical equilibrium, as follows from \eqref{composition}, chemical imbalances of all the considered reactions vanish, $\mu_{\rm in\leftrightarrow out \, 0}=0$, and, therefore, direct and inverse reactions proceed (if allowed) at the coinciding rates. 

Further, regarding {\it hydrostatic equilibrium}, this study formally treats a neutron star as rotating {\it slowly} with angular velocity $\Omega\ll \Omega_{\rm K}$, where $\Omega_{\rm K}=\sqrt{GM/R^3}$ is the Keplerian angular velocity. In fact, for $r$-modes, the actual small parameter involved in the equations is $(\Omega/\Omega_{\rm K})^2$, which makes the slow rotation approximation valid for application to most LMXBs, while still providing adequate estimates even for those with the highest rotation rates, peaking at $\Omega\simeq 0.5 \, \Omega_{\rm K}$ (see, however, the discussion of $r$-mode nonanalyticity in Sec.\,\ref{Nonanalytic r-modes}). Thus, in what follows, we describe hydrostatic equilibrium using the metric tensor $\rmg_{\mu\nu}$ of a slowly rotating neutron star, which in the spherical coordinates $x^\mu\equiv(ct,r,\theta,\varphi)$ is given by \cite{hartle1967}%
\footnote{
Note that our notations slightly differ from those used in the original paper \cite{hartle1967} by Hartle: $\nu=\nu_{\rm Hartle}/2$ and $\lambda=\lambda_{\rm Hartle}/2$.
}
%
\begin{multline}
\label{geometry}
ds^2\equiv{\rm g}_{\mu\nu}dx^\mu dx^\nu=-e^{2\nu(r)}c^2 dt^2+e^{2\lambda(r)}dr^2+\\
+r^2\{d\theta^2+\sin^2\theta[d\varphi-\Omega\omega(r)dt]^2\}+O(\Omega^2/\Omega_{\rm K}^2).
\end{multline}
The equilibrium four velocity $u^\mu\equiv\fru^\mu_0$, in turn, can be written in the chosen coordinates as
\begin{gather}
\label{velocity}
u^\mu=\Lambda k^\mu, \quad k^\mu=\left(1,0,0,\frac{\Omega}{c}\right)=\delta^\mu_{t}+\frac{\Omega}{c}\delta^\mu_{\varphi}, \\
\label{redshift}
\Lambda=(-k^\mu k_\mu)^{-1/2}=e^{-\nu(r)}+O(\Omega^2/\Omega_{\rm K}^2),
\end{gather}
where we introduced the gravitational redshift $\Lambda$. One can verify that the vector $k^\mu$ satisfies the Killing equation: $\nabla_\mu k_\rho+\nabla_\rho k_\mu=0$. Using this equation, equations \eqref{velocity}-\eqref{redshift}, and equality $u^\rho\nabla_\rho\Lambda=0$, it is straightforward to show that 
\begin{gather}
\label{DuEquilibrium}
u^\rho\nabla_\rho u_\mu=-\frac{1}{\Lambda}\nabla_\mu\Lambda.
\end{gather}
For simplicity, in our calculations, we will ignore the oblateness of the neutron star due to rotation and discard the $O(\Omega^2/\Omega_{\rm K}^2)$ terms in \eqref{geometry} and \eqref{redshift}. This approximation allows us to treat $p_0(r)$, $\varepsilon_0(r)$, $n_{b0}(r)$ and other thermodynamic quantities as $\theta$-independent functions of a single variable $r$. These functions, as well as metric coefficients $\nu(r)$ and $\lambda(r)$, can be determined by solving the well-known Tolman-Oppenheimer-Volkoff (TOV) equations. The function $\omega(r)$, which characterizes the frame-dragging effect mentioned in the introduction, in turn, can be found by solving Hartle's equation \cite{hartle1967}.

Finally, {\it thermal equlibrium} in GR corresponds to the constant {\it redshifted} (i.e., accounting for the gravitational redshift) {\it temperature} $T^\infty$ \cite{mirallesetal1993},
\begin{gather}
\label{thermal equilibrium}
T^\infty=\const, \qquad T^\infty\equiv \frac{T}{\Lambda}\approx T(r) e^{\nu(r)},
\end{gather}
where $T(r)$ is the {\it local temperature}, measured at a given point $r$ by a local observer in a reference frame, rotating with angular velocity $\Omega$.


\subsection{Neutron star oscillation equations}\label{NS oscillations}


\subsubsection{General equations}\label{General equations}

In this section, we consider linear periodic perturbations over the described neutron star equilibrium. Specifically, we study perturbations of the form $\delta f\propto e^{i\sigma t+i m \varphi}$, where $\delta f\equiv f-f_0$ is the Eulerian perturbation of a quantity $f$, $\sigma$ is the oscillation frequency, measured by a distant inertial observer, and $m$ is the azimuthal quantum number. We perform our modeling in the Cowling approximation; i.e., we ignore perturbations $\delta\frg_{\mu\nu}$ of the gravitational field in hydrodynamic equations and assume that the metric tensor therein coincides with the equilibrium one, given by \eqref{geometry}. This approximation provides reasonable estimates of the NS oscillation properties  \cite{lsr1990, yk1997, jc2017} (see also \cite{kgk2022_1, kgk2022_2} for discussion regarding relativistic $r$-modes) and, at the same time, allows for their hydrodynamic modeling instead of a more complex one, based on Einstein equations.

We describe the NS oscillations using the Lagrangian displacement $\xi^{\mu}$, that shows the induced variations of the fluid element world lines. In the Cowling approximation, $\xi^\mu$ is related to the velocity perturbation as%
\footnote{
This formula follows from the general relation (see, e.g., \cite{friedman1978}) between velocity perturbation and Lagrangian displacement: in the Cowling approximation, $\delta\fru^\mu=\perp^\mu_\rho\mathcal{L}_u\xi^\rho$, where $\mathcal{L}_u$ is the Lie derivative along $u^\mu$. To derive it, one has to substitute the equilibrium velocity \eqref{velocity}, use the conventional gauge condition $u^\mu \xi_\mu=0$ and orthogonality $u_\mu \perp^\mu_\rho=0$.
}
%
\begin{gather}
\label{Lagrangian displacement}
\delta\fru^\mu=i(\sigma_{\rm loc}/c)\xi^\mu, \qquad \sigma_{\rm loc}=\Lambda(\sigma+m\Omega),
\end{gather}
where $\sigma_{\rm loc}(r)$ is the (varying throughout the star) {\it local oscillation frequency}, measured by a local observer in the reference frame, rotating with angular velocity $\Omega$. Note that Lagrangian displacement $\xi^\mu$ satisfies the gauge condition, $u_\mu \xi^\mu=0$, that ensures the lineraized velocity normalization condition: $\delta(\fru^\mu\fru_\mu)=2u_\mu\delta\fru^\mu=0$.

For scalar quantities $f$, we find convenient to use Lagrangian perturbations $\Delta f$, defined as%
\footnote{
Generally, for a tensor quantity of arbitrary rank, Lagrangian perturbation can be defined as $\Delta\equiv\delta+\mathcal{L}_\xi$, where $\mathcal{L}_\xi$ is the Lie derivative along the vector $\xi^\mu$ (see, e.g., \cite{friedman1978}). For scalars, this definition reduces to the one provided in the text.
}
%
\begin{gather}
\Delta f\equiv \delta f+\xi^\mu \nabla_\mu f_0.
\end{gather}
In these notations, the system of equations that govern linear oscillations can be written in the following form:
\begin{gather}
\label{system}
\left\{
\begin{gathered}
\sigma_{\rm loc}^2\xi^\mu-2i c \sigma_{\rm loc}\xi^\rho\nabla_\rho u^\mu=\frac{c^2}{w_0}\biggl[\pnabla^\mu\delta p-\frac{\delta w}{w_0}\pnabla^\mu p_0\biggr]+\\
\hfill+\text{[viscous terms containing $\eta$]} \\
i\sigma_{\rm loc}[\Delta n_{k}+n_{k0}\pnabla_\mu \xi^\mu]=\Gamma_k \hfill \\
\delta p=\biggl(\frac{\p p}{\p n_{k}}\biggr)_0\delta n_{k}, \quad \delta\varepsilon=\mu_{k0}\delta n_k, \quad \delta w=\delta p+\delta\varepsilon. \hfill
\end{gathered}
\right.\raisetag{1.4cm}
\end{gather}
Here, the first equation is the linearized Euler equation \eqref{EulerEq}, the second is the set of linearized continuity equations \eqref{ContEq}, and the remaining equations follow from the EOS \eqref{EOS} and thermodynamic relations \eqref{thermodynamics}.

In practice, unless one is particularly interested in number density perturbations, the continuity equations are inconvenient to work with. Therefore, in what follows, we will manipulate them to obtain two equations for the energy density and pressure perturbations. We {\it announce} that, generally, the result can be written in the following form
\begin{gather}
\label{Delta epsilon eq}
\Delta\varepsilon+w_0\pnabla_\mu\xi^\mu=0, \\
\label{Delta p equation}
\Delta p+\gamma p_0\pnabla_\mu\xi^\mu=-\zeta c \nabla_\mu \delta \fru^\mu,
\end{gather}
where $\gamma$ and $\zeta$ are the previously mentioned {\it adiabatic index} and {\it bulk viscosity coefficient}, respectively. As will be shown shortly, chemical reactions not only determine the magnitude of bulk viscosity $\zeta$, but they also modify the adiabatic index $\gamma$.


\subsubsection{Sources powered by chemical reactions}

In order to derive the announced equations for $\Delta\varepsilon$ \eqref{Delta epsilon eq} and $\Delta p$ \eqref{Delta p equation}, as well as to determine bulk viscosity $\zeta$ and adiabatic index $\gamma$, we must specify the sources $\Gamma_k$ in linearized continuity equations \eqref{system}. As mentioned earlier, these sources are powered by chemical reactions, driven out of equilibrium in the course of oscillations. Previously  (see Sec.\,\ref{Neutron star equilibrium}), we divided relevant chemical reactions into three groups: strong process \eqref{sp}, weak nonleptonic processes \eqref{wnpFirst}-\eqref{wnpLast}, and weak leptonic processes, such as \eqref{wlpFirst}-\eqref{wlpLast}. Each of these groups has distinct physical properties and will be treated accordingly. 

{\it First, let us consider the strong process  \eqref{sp}}. When protons are normal, this process can be safely treated as equilibrated, since strong interactions bring it to equilibrium on timescales much shorter than hydrodynamic oscillation period. For superconducting protons, however, this may not hold, as proton superconductivity strongly inhibits the reaction at sufficiently low temperatures. We will return to this issue in Sec.\,\ref{SfSc effects}. Here, for simplicity, we will formally treat the strong process as being in equilibrium even when protons are superconducting. 

At least for our microphysical input (see Sec.\,\ref{Numerical results}), the strong process ignites as soon as $\Xi^-$-hyperons appear, which (within the approximations made) implies that their chemical potential satisfies
\begin{gather}
\label{strong equilibrium}
\mu_{\Xi^-}=2\mu_\Lambda-\mu_{\rm p}
\end{gather}
with high precision [see the condition of chemical equilibrium \eqref{chemical equilibrium}]. This condition, combined with quasineutrality of the matter, $e_k n_k=0$, allows to reduce the number of independent thermodynamic variables from six $\{n_{\rm n},n_{\rm p},n_{\rm e},n_\mu,n_\Lambda,n_{\Xi^-}\}$ to four. In this study, we follow closely \cite{ofengeimetal2019} and work in the $\{n_b,y_{\rm e},y_\mu,y_{\rm s}\}$ variables, defined according to \eqref{new variables}. When using these variables, it is sufficient to consider only the baryon source $\Gamma_b \equiv b_k \Gamma_k$, lepton sources $\Gamma_{\rm e, \mu}$, and the strangeness source $\Gamma_{\rm s} \equiv s_k \Gamma_k$. In fact, since baryon number is conserved in all chemical reactions, baryon source $\Gamma_b=0$ vanishes, which leaves us with $\Gamma_{\rm e,\mu,s}$ being the only relevant sources. It is also worth noting that, in the chosen variables, we have
\begin{gather}
\label{energy derivatives}
\biggl(\frac{\p\varepsilon}{\p n_b}\biggr)_0=\frac{w_0}{n_{b0}}, \qquad \biggl(\frac{\p\varepsilon}{\p y_{\rm s}}\biggr)_0=0,
\end{gather}
and Lagrangian perturbation $\Delta f$ of any thermodynamic quantity $f$ can be written as
\begin{multline}
\label{Delta f}
\Delta f=\biggl(\frac{\p f}{\p n_b}\biggr)_0 \Delta n_b+\biggl(\frac{\p f}{\p y_{\rm s}}\biggr)_0 \Delta y_{\rm s}+\\
+\biggl(\frac{\p f}{\p y_{\rm e}}\biggr)_0 \Delta y_{\rm e}+\biggl(\frac{\p f}{\p y_{\rm \mu}}\biggr)_0 \Delta y_{\rm \mu}.
\end{multline}

{\it Next, weak leptonic and nonleptonic reactions}, being mediated by weak interaction, are substantially slower than the strong process and, generally, their deviation from equilibrium should be accounted for. Weak leptonic reactions are typically much slower than weak nonleptonic ones, and, in what follows, we ignore their contributions to the sources $\Gamma_k$. Physically this means that, although driven out of equilibrium, these reactions are too slow to produce sizable contribution to $\Gamma_k$. In particular, since these are the only processes that change lepton numbers, this means that leptons are approximately conserved: $\Gamma_{\rm e}=\Gamma_{\mu}=0$.

{\it Finally}, we treat {\it weak nonleptonic processes} \eqref{wnpFirst}-\eqref{wnpLast} as arbitrarily fast, but staying within the subthermal limit. The latter means that, the oscillation amplitude is sufficiently small, so that, for a given pair $(12\leftrightarrow 34)$ of direct $(\rightarrow)$ and inverse $(\leftarrow)$ reactions, the corresponding chemical imbalance
\begin{gather}
\mu_{12\leftrightarrow 34}\equiv \mu_1+\mu_2-\mu_3-\mu_4
\end{gather}
is small: $|\mu_{12\leftrightarrow 34}|\ll k_{\rm B}T$. In this limit, the difference $\Gamma_{12\leftrightarrow 34}$ in the rates (i.e., the number of reactions per unit time per unit volume) of direct and inverse processes, induced by an oscillation, can be Taylor-expanded as
\begin{gather}
\label{lambdas0}
\Gamma_{12\leftrightarrow 34}\equiv\lambda_{12\leftrightarrow 34}\mu_{12\leftrightarrow 34}=\lambda_{12\leftrightarrow 34}\Delta\mu_{12\leftrightarrow 34},
\end{gather}
where we used that chemical imbalances vanish in equilibrium, $\mu_{12\leftrightarrow 34 \, 0}=0$, and, therefore, $\mu_{12\leftrightarrow 34}=\Delta\mu_{12\leftrightarrow 34}$. It is easy to see that, due to equilibrium condition \eqref{strong equilibrium}, all weak nonleptonic reactions possess the shared chemical imbalance
\begin{gather}
\mu_{12\leftrightarrow 34}=\mu\equiv \mu_{\rm n}-\mu_\Lambda,
\end{gather}
and the equation \eqref{lambdas0} further reduces to
\begin{gather}
\label{lambdas}
\Gamma_{12\leftrightarrow 34}=\lambda_{12\leftrightarrow 34}\Delta\mu.
\end{gather}
In the literature, the quantities $\lambda_{12\leftrightarrow 34}$ (as well as $\Gamma_{12\leftrightarrow 34}$) are often referred to as {\it reaction rates}. Their explicit values are provided by microphysical calculations.

Summarizing, in our approach, the strangeness source $\Gamma_{\rm s}$ is the only nonvanishing source. Meanwhile, of all considered reactions, only weak nonleptonic processes \eqref{wnpFirst}-\eqref{wnpLast} contribute to $\Gamma_{\rm s}$, with each (direct) process changing strangeness by $(-1)$. Keeping this in mind, we present the resulting expressions for the sources of interest as follows:
\begin{gather}
\label{sources}
\begin{gathered}
\Gamma_b=\Gamma_{\rm e}=\Gamma_\mu=0, \\
\Gamma_{\rm s}=-\lambda_{\rm tot}\Delta\mu, \quad
\lambda_{\rm tot}\equiv\sum_{12\leftrightarrow 34}\lambda_{12\leftrightarrow 34},
\end{gathered}
\end{gather}
where summation is performed over all weak nonleptonic processes, and $\lambda_{\rm tot}$ is their ``total" reaction rate.


\subsubsection{Bulk viscosity and adiabatic index}\label{Bulk viscosity and adiabatic index}

Now, we are in position to derive the announced equations \eqref{Delta epsilon eq} and \eqref{Delta p equation} for $\Delta\varepsilon$ and $\Delta p$. First, using linearized continuity equations \eqref{system}, definitions \eqref{new variables}, and explicit form of sources \eqref{sources}, we find
\begin{gather}
\label{ContEqNewVar}
\left\{
\begin{gathered}
\Delta n_b+n_{b0}\pnabla_\mu\xi^\mu=0 \hfill \\
\Delta y_{\rm e}=\Delta y_{\mu}=0 \hfill \\
\Delta y_{\rm s}=i(\lambda_{\rm tot}/n_{b0} \sigma_{\rm loc})\Delta\mu. \hfill
\end{gathered}
\right.
\end{gather}
Then, we use \eqref{Delta f} to exclude the chemical imbalance perturbation $\Delta\mu$ from this system, which allows further rewriting the only nonvanishing Lagrangian perturbations of the independent thermodynamic variables as
\begin{gather}
\begin{gathered}
\label{Deltas}
\Delta n_{b}=-n_{b0}\pnabla_\mu\xi^\mu, \\
\Delta y_{\rm s}=-i\frac{(\lambda_{\rm tot}/\lambda_{\rm max})}{1-i(\lambda_{\rm tot}/\lambda_{\rm max})}\frac{(\p\mu/\p n_b)_0}{(\p \mu/\p y_{\rm s})_0}n_{b0}\pnabla_\mu\xi^\mu,
\end{gathered}
\end{gather}
where we introduced 
\begin{gather}
\lambda_{\rm max}\equiv n_{b0}\sigma_{\rm loc}(\p\mu/\p y_{\rm s})_0^{-1}.
\end{gather}

Finally, to derive the desired expressions for $\Delta \varepsilon$ and $\Delta p$, we apply the general formula \eqref{Delta f} for Lagrangian perturbations and use the expressions \eqref{Deltas} for $\Delta n_b$ and $\Delta y_{\rm s}$, recalling that $\Delta y_{\rm e}=\Delta y_{\rm \mu}=0$ and that energy density derivatives satisfy \eqref{energy derivatives}. Then, as a final step, we note that
\begin{gather}
\label{divXi relation}
\pnabla_\mu\xi^\mu=(1/\Lambda)\nabla_\mu(\Lambda\xi^\mu)=-(i/\sigma_{\rm loc})c\nabla_\mu\delta u^\mu. 
\end{gather}
The first equality follows directly from the gauge condition, $u_\mu\xi^\mu=0$, and the equation \eqref{DuEquilibrium}, while the second equality follows trivially from the equation \eqref{Lagrangian displacement}. Using \eqref{divXi relation} to ``exclude" (wherever necessary) the imaginary unit $i$, we obtain the announced equations \eqref{Delta epsilon eq} and \eqref{Delta p equation} with
\begin{gather}
\label{gamma}
\gamma=\frac{n_{b0}}{p_0}\biggl(\frac{\p p}{\p n_b}\biggr)_0\biggl[1-\beta\frac{(\lambda_{\rm tot}/\lambda_{\rm max})^2}{1+(\lambda_{\rm tot}/\lambda_{\rm max})^2}\biggr], \\
\label{zeta}
\zeta=\zeta_{\rm max}\frac{2(\lambda_{\rm tot}/\lambda_{\rm max})}{1+(\lambda_{\rm tot}/\lambda_{\rm max})^2},
\end{gather}
and functions $\zeta_{\rm max}$ and $\beta$, defined as
\begin{gather}
\hspace{-0.1cm}\zeta_{\rm max}\equiv\frac{n_{b0}\beta}{2\sigma_{\rm loc}}\biggl(\frac{\p p}{\p n_b}\biggr)_0, \quad \hspace{-0.2cm} \beta\equiv\frac{(\p p/\p y_{\rm s})_0}{(\p p/\p n_b)_0}\frac{(\p\mu/\p n_b)_0}{(\p\mu/\p y_{\rm s})_0}.
\end{gather}

The parameters $\lambda_{\rm max}$, $\zeta_{\rm max}$, and $\beta$ have transparent physical interpretation. It is evident that, as a function of reaction rate $\lambda_{\rm tot}$, bulk viscosity reaches its maximum value $\zeta=\zeta_{\rm max}$ at $\lambda_{\rm tot}=\lambda_{\rm max}$, and vanishes far away from this point, where reactions are either too slow ($\lambda_{\rm tot}\ll\lambda_{\rm  max}$) or too fast ($\lambda_{\rm tot}\gg\lambda_{\rm  max}$)%
\footnote{
Note that, strictly speaking, oscillation frequency $\sigma$ and, therefore, local frequency $\sigma_{\rm loc}(r)$ generally can take complex values due to dissipation in the system. In NS conditions, however, the dissipation is weak and the associated imaginary part of the frequency is negligibly small, which allows its inclusion via perturbative approach, as discussed in Sec.\,\ref{Instability windows}. In the leading order of this approach, we can ignore the imaginary part of $\sigma$ and, therefore, consider $\lambda_{\rm max}$ and $\zeta_{\rm max}$ as real-valued.
}.
The parameter $\beta$, in turn, describes the difference between the values of the adiabatic index in these two limiting regimes. These are well-known and expected results, as clarified below.

Let us first consider the case when weak nonleptonic reactions are too slow ($\lambda_{\rm tot}\ll\lambda_{\rm  max}$). In this limit, the matter lacks the time to respond to the changes (compression/rarefaction), induced by oscillations, and its chemical composition remains the same as if there were no weak reactions at all: $\Delta y_{\rm s,e,\mu}=0$. This explains vanishing bulk viscosity and leads to the adiabatic index
\begin{gather}
\label{gamma frozen}
\gamma_{\rm frozen}\equiv \frac{n_{b0}}{p_0}\biggl(\frac{\p p}{\p n_b}\biggr)_0.
\end{gather}
Here and below, we use the term {\it``frozen"} to indicate that weak nonleptonic reactions are negligibly slow and thus have no effect on the adiabatic index. Note, however, that since the strong process \eqref{sp} is still accounted for, this limit does not correspond to a fully frozen composition of the matter ($\Delta y_k = 0$ for all particle species) with {\it all} reactions being switched off.

Now, let us consider the opposite case, when weak nonleptonic reactions are extremely fast ($\lambda_{\rm tot}\gg \lambda_{\rm max}$). In this limit, the matter immediately responds to oscillations and, therefore, maintains chemical equilibrium, which leads to vanishing bulk viscosity. At the same time, fast chemical reactions modify the adiabatic index, that in the considered limit takes the value
\begin{gather}
\label{gamma fast}
\gamma_{\rm fast}\equiv\gamma_{\rm frozen}(1-\beta).
\end{gather}
Here and below, we use the term {\it ``fast"} to indicate that, when calculating the adiabatic index, weak nonleptonic reactions are treated as extremely fast. As announced, according to equation \eqref{gamma fast}, the coefficient $\beta$ describes the relative correction to the frozen adiabatic index arising in the case of extremely fast chemical reactions.

\begin{figure}
\centering
\includegraphics[width=1.0\linewidth]{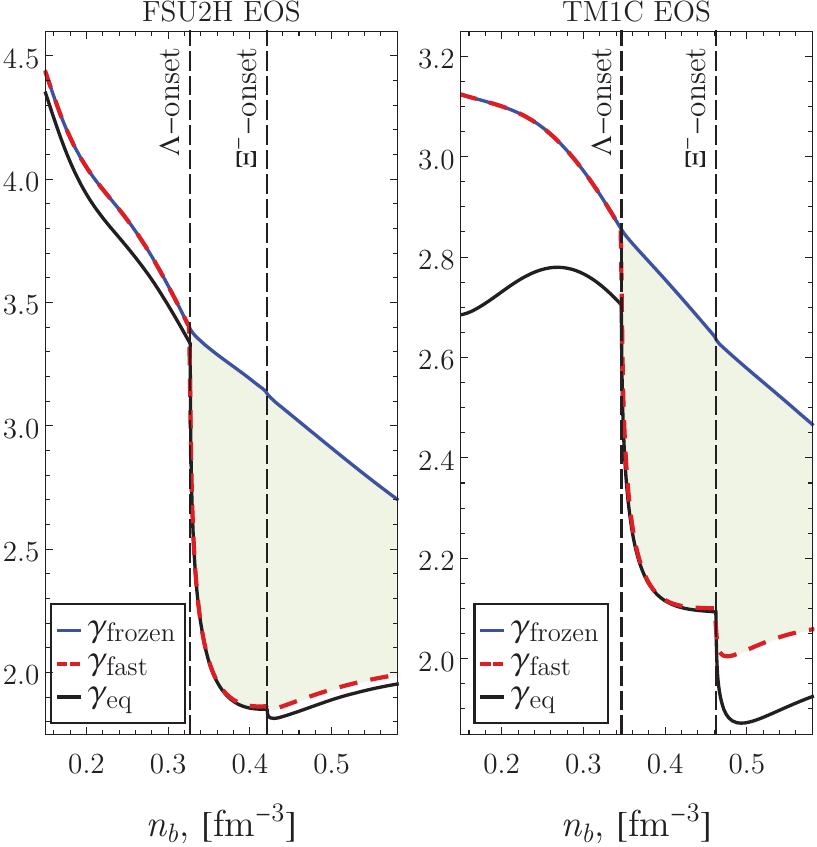}
\caption{Adiabatic index as a function of baryon number density $n_b$ for FSU2H EOS (left panel) and TM1C EOS (right panel). Blue lines (``frozen") show the limit corresponding to negligibly slow weak nonleptonic reactions [but still accounting for the strong process \eqref{sp}]. Red dashes (``fast") show the opposite limit when all considered weak nonleptonic reactions are assumed to be extremely fast. Black lines (``eq") show the adiabatic index calculated assuming full thermodynamic equilibrium. Depending on physical conditions, the actual adiabatic index (in the case of equilibrated strong process) takes values inside the green shaded area.}
\label{adiabatic fig}
\end{figure}

Finally, as follows from \eqref{gamma}, the actual value $\gamma$ of the adiabatic index lies in the range $\gamma_{\rm fast}\leq\gamma\leq\gamma_{\rm frozen}$. This behavior is illustrated by Fig.\,\ref{adiabatic fig}, where we plot $\gamma_{\rm frozen}$ (blue line) and $\gamma_{\rm fast}$ (red dashes) as functions of the baryon number density, while the shaded green region shows the possible values of the actual adiabatic index $\gamma$ given by \eqref{gamma} [see Sec.\,\ref{Numerical results} for details on microphysical input]. The figure also shows the so-called equilibrium adiabatic index $\gamma_{\rm eq}$  \cite{gck2014}, calculated assuming the full thermodynamic equilibrium of the matter, where all thermodynamic functions can be considered as depending solely on the baryon number density (recall Sec.\,\ref{Neutron star equilibrium}),
\begin{gather}
\label{gamma eq}
\gamma_{\rm eq}\equiv\frac{n_{b0}}{p_0}\frac{dp_0}{d n_{b0}}.
\end{gather}
%

\subsection{Estimating the effects of neutron superfluidity and proton superconductivity}\label{SfSc effects}

In this section, we briefly discuss how neutron superfluidity and proton superconductivity affect NS oscillations. For simplicity, we estimate their influence by the corresponding modifications to the shear viscosity $\eta$ and the reaction rates, while neglecting the changes they bring to the form of hydrodynamic equations. As we mentioned previously, we allow protons to be either normal or strongly superconducting, while neutron superfluidity, depending on the temperature, is allowed to be arbitrarily strong. 

Shear viscosity quantifies the friction in the fluid arising due to velocity gradients. In nucleonic NS interiors, this friction primarily arises from collisions of leptons with other charged particle species, mediated by electromagnetic interaction \cite{ss2018}. This picture is likely preserved with appearance of neutral $\Lambda$-hyperons. The presence of charged $\Xi^-$-hyperons, in turn, may reduce (due to quasineutrality) the lepton fraction, thereby suppressing lepton contribution to $\eta$. Similar effect takes place in $\rm npe\mu\Sigma^-\Lambda$-matter, where neutrons commence to provide the dominant contribution to the shear viscosity at densities with sufficiently high fraction of $\Sigma^-$-hyperons \cite{sv2021}. Thus, for $\rm npe\mu\Lambda\Xi^-$-matter, we expect the influence of neutron superfluidity on shear viscosity to be limited to regions containing sizable fraction of $\Xi^-$-hyperons (although detailed study may be required to confirm this). None of the NS models considered in this work (see Sec.~\ref{Numerical results}) contain such regions, implying that $\eta$ remains only weakly influenced by neutron superfluidity in these cases. Proton superconductivity, in turn, can substantially modify shear viscosity, as it suppresses lepton-proton scattering and modifies the screening properties of the matter (see review \cite{ss2018} and references therein). In this study, aimed at application to the $r$-mode instability windows in hyperonic neutron stars, a precise consideration of shear viscosity is unnecessary, as its role in this context is negligible compared to that of hyperonic bulk viscosity. For this reason, here we estimate $\eta$ and its modifications by proton superconductivity using the relevant expressions in nucleonic matter \cite{ss2018}.

Bulk viscosity $\zeta$ and adiabatic index $\gamma$ are affected by nucleon superfluidity and superconductivity through reaction rates of the chemical reactions under consideration. Here, we model strong proton superconductivity by ``switching off" the weak nonleptonic reactions \eqref{npLp} and \eqref{wnpLast} that involve protons, which is achieved by setting the corresponding reaction rates $\lambda_{12\leftrightarrow 34}$ to zero. Strictly speaking, depending on the temperature, proton supercoductivity may also suppress the strong process \eqref{sp}. We will estimate the relevant temperature scales at the end of this section. For now, however, we assume for simplicity that proton superconductivity is not sufficiently strong to suppress the strong process, which therefore remains equilibrated. Concerning neutron superfluidity, to describe its effect on reaction rates we follow closely \cite{haenseletal2000,haenseletal2001, haenseletal2002} and introduce the so-called {\it reduction factors},
\begin{gather}
\label{ReductionFactors}
\mathcal{R}_{12\leftrightarrow 34}\equiv \lambda_{12\leftrightarrow 34}^{\rm superfluid}/\lambda_{12\leftrightarrow 34}^{\rm normal}, 
\end{gather}
showing the difference of rates $\lambda_{12\leftrightarrow 34}^{\rm normal}$ in normal matter and those $\lambda_{12\leftrightarrow 34}^{\rm superfluid}$ in matter with superfluid neutrons.

The difference arises from the changes that neutron superfluidity brings to the neutron energy spectrum. Generally, in the vicinity of the Fermi surface  $|\bs{p}|=p_{{\rm F}k}$, the energy spectrum $\varepsilon_k(\bs{p})$ of a normal particle species $k$ (including neutrons, if they are normal) takes the form
\begin{gather}
\varepsilon_k(\bs{p})=\mu_k + v_{{\rm F}k}(|\bs{p}|-p_{{\rm F}k}),
\end{gather}
where ${\bs p}$ is the particle momentum, $p_{{\rm F}k}$ is the Fermi momentum, and $v_{{\rm F}k}$ is the Fermi velocity of the specified particle species. Now, if we let neutrons be superfluid, which happens at temperatures below the {\it critical temperature} $T_{\rm c}$, their spectrum acquires the energy gap $\delta$ (e.g., \cite{lp80,haenseletal2000,haenseletal2001, haenseletal2002}),
\begin{gather}
\label{nSpectrum}
\hspace{-0.12cm}\varepsilon_{\rm n}({\bs p})=\mu_{\rm n}+\sgn(|\bs{p}|-p_{{\rm F}{\rm n}})\sqrt{\delta^2+v_{{\rm F}{\rm n}}^2(|\bs{p}|-p_{{\rm F}{\rm n}})^2}. \raisetag{0.5cm}
\end{gather}
The explicit form of the gap is determined by the microphysical properties of the interaction, responsible for the Cooper pair formation. Generally, it can depend solely on the temperature (isotropic gap), or it may also depend on the angle $\vartheta$ between the momentum ${\bs p}$ and the quantization axis (anisotropic gap). Neutrons in the NS core, considered here, experience the triplet-state pairing, resulting in an anisotropic gap%
\footnote{This particular gap corresponds to a case of triplet-state pairing, when the total angular momentum projection $m_J$ of the Cooper pair onto the quantization axis vanishes, $m_J=0$.},
%
\begin{gather}
\delta(T,\vartheta)=\Delta(T)\sqrt{1+3 \cos^2\vartheta}. \label{gap}
\end{gather}
At temperatures $T\leq T_{\rm c}$ below that of superfluidity onset, the temperature dependence $\Delta(T)$ of the gap in the case of the triplet-state pairing is given by (see, e.g., \cite{yls99})
\begin{gather}
\label{DeltaFromT}
\frac{\Delta(T)}{k_{\rm B} T}=\sqrt{1-\frac{T}{T_{\rm c}}}\biggl[0.7893+\frac{1.188}{T/T_{\rm c}}\biggr].
\end{gather}

The precise consideration of anisotropic pairing is rather complicated and here, for simplicity, we follow the approximate approach of \cite{bhy2001} and replace the anisotropic gap $\delta(T,\vartheta)$ in Eq.\ \eqref{nSpectrum} by the {\it effective isotropic gap}
\begin{gather}
\delta_{\rm eff}(T)\equiv\min\{\delta(T,\vartheta)|_{|\bs{p}|=p_{\rm Fn}}\}=\Delta(T),
\end{gather}
assuming that the latter, nevertheless, obeys the temperature dependence \eqref{DeltaFromT} (see also \cite{gh2005, leinson17, gg2023} for a discussion and application of this approximation).

Within this simplified approach, the reaction rate $\Gamma_{12\leftrightarrow 34}$ for degenerate matter can be split into a product of angular and energy integrals (see technical details, e.g., in \cite{haenseletal2000, haenseletal2001, ofengeimetal2019}): the angular integral is common to both the superfluid and the normal cases, whereas it is the energy integral that is modified by neutron superfluidity. In normal matter, adopting the general expression for reaction rates from \cite{ofengeimetal2019} and omitting the angular integral, we have
\begin{gather}
\Gamma_{12\leftrightarrow 34}^{\rm normal}\propto \biggl[\mathcal{I}\biggl(\frac{\mu}{k_{\rm B}T}\biggr)-\mathcal{I}\biggl(-\frac{\mu}{k_{\rm B}T}\biggr)\biggr]\propto \mathcal{I}'(0)\mu \label{dGammaNormal}
\end{gather}
with the function $\mathcal{I}(v)$ defined by
\begin{gather}
\mathcal{I}(v)\equiv\int \delta\biggl(\sum_{m=1}^4 x_m-v\biggr) \prod_{k=1}^4\frac{dx_k}{1+e^{x_k}}. \label{Inormal}
\end{gather}
Here, $\delta(x)$ denotes the Dirac delta function (not to be confused with the Eulerian perturbation). The physical origin of the factor \eqref{dGammaNormal} is simple. The energies $\varepsilon_k$ of particle species $k$ are distributed according to Fermi-Dirac function $f_k=1/[1+e^{(\varepsilon_k-\mu_k)/k_{\rm B}T}]$, that for normal particle species reduces to
\begin{gather}
f_k=\frac{1}{1+e^{x_k}}, \qquad x_k\equiv \frac{v_{{\rm F}k}(|\bs{p}|-p_{{\rm F}k})}{k_{\rm B} T}.
\end{gather}
The integral \eqref{Inormal} then arises as a result of summation over all particle conversions $12\leftrightarrow 34$, allowed by the energy conservation law. Now, if we let neutrons be superfluid, their energy spectrum acquires the energy gap $\delta_{\rm eff}$, which, according to \eqref{nSpectrum}, can be accounted for by replacing the neutron variable
\begin{gather}
x_{\rm n}\to\sgn(x_{\rm n})\sqrt{y^2+x_{\rm n}^2}, \qquad y\equiv\delta_{\rm eff}/k_{\rm B}T.
\end{gather}
This modification is conveniently described by introducing new variables $z_k$, defined as follows: $z_k=x_k$ for normal particles, and $z_k=\sgn(x_k)\sqrt{y^2+x_k^2}$ for superfluid neutrons. Using this notation, in the case of superfluid neutrons, the reaction rates can be written as
\begin{gather}
\Gamma_{12\leftrightarrow 34}^{\rm superfluid}\propto \mathcal{I}^{[n]}{}'(0)\mu, \label{dGammaSuperfluid}
\end{gather}
where
\begin{gather}
\mathcal{I}^{[n]}(v)\equiv\int \delta\biggl(\sum_{m=1}^4 z_m-v\biggr)\prod_{k=1}^4\frac{dx_k}{1+e^{z_k}}. \label{Isuperfluid}
\end{gather}
Here, we use the superscript $[n]$ to stress that the value of $\mathcal{I}^{[n]}(v)$ depends on the number $[n]$ of neutrons, involved in the process under consideration.

\begin{figure}
\centering
\includegraphics[width=1.0\linewidth]{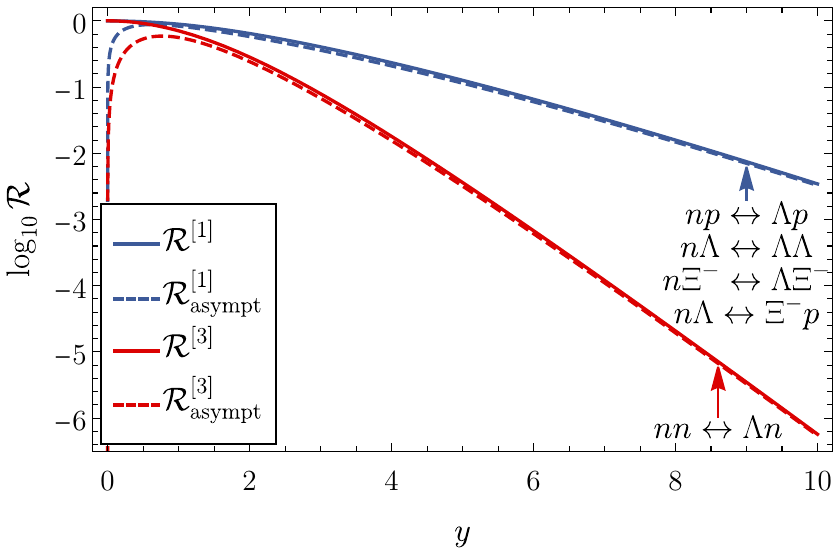}
\vspace{-0.6cm}
\caption{Reduction factors for processes involving either one (shown in blue) or three (shown in red) superfluid neutrons. Solid lines show the exact values, obtained via numerical integration, while dashes show the asymptotic formulas, obtained assuming large values of $y$.}
\label{reduction_factors_pic}
\end{figure}

Now, reduction factors are easily found using \eqref{lambdas}, \eqref{ReductionFactors} and \eqref{dGammaNormal}-\eqref{Isuperfluid}. Considered reactions \eqref{wnpFirst}-\eqref{wnpLast} involve either one or three neutrons, and the effect of neutron superfluidity on their reaction rates is described by the corresponding reduction factors $\mathcal{R}^{[1]}$ and $\mathcal{R}^{[3]}$,
\begin{gather}
\mathcal{R}^{[1]}=\frac{3}{\pi^2}\mathcal{I}^{[1]}{}'(0), \qquad \mathcal{R}^{[3]}=\frac{3}{\pi^2}\mathcal{I}^{[3]}{}'(0),
\end{gather}
where we used that $\mathcal{I}'(0)=\pi^2/3$ (see, e.g, \cite{ofengeimetal2019}). These reduction factors depend solely on the dimensionless energy gap $y$. One can show that, at large values of $y$, they exhibit the following asymptotic behavior:
\begin{gather}
\label{R1}
\mathcal{R}^{[1]}_{\rm asympt}=\frac{3}{2}\sqrt{\frac{\pi y}{2}}e^{-y} \frac{\left(y^2+y+\pi ^2\right)}{\pi ^{2}}, \\
\label{R3}
\mathcal{R}^{[3]}_{\rm asympt}=\frac{3 \sqrt{3} }{2 \pi }e^{-2 y} \, y \, (3 \, y+2).
\end{gather}
Here, the expression \eqref{R1} was adopted from \cite{haenseletal2002}, while the formula \eqref{R3} represents an original result that, as far as we know, has not been obtained elsewhere. These expressions indicate that, as temperature decreases (i.e., as $y$ increases), superfluidity eventually leads to the exponential suppression of reaction rates. We confirm this behavior by straightforward numerical calculation, as demonstrated in Fig.\,\ref{reduction_factors_pic}. This figure displays the exact (solid lines) and asymptotic (dashed lines) reduction factors $\mathcal{R}^{[1]}$ (shown in blue) and $\mathcal{R}^{[3]}$ (shown in red). We also accompany each curve with the corresponding weak nonleptonic reactions. 

According to \eqref{gamma} and \eqref{zeta}, these results imply that, at temperatures $T<T_{\rm c}$, sufficiently below that of neutron superfluidity onset, the bulk viscosity $\zeta$ is exponentially suppressed, while the deviation of the adiabatic index from the frozen one \eqref{gamma frozen} is negligibly small. At temperatures above $T_{\rm c}$, however, it is hard to predict whether the matter is in the frozen regime or not. 

For further applications, it is beneficial to have simplified expressions for the reduction factors, valid over a broad range of $y$.  According to \cite{haenseletal2002}, the factor $\mathcal{R}^{[1]}$ is with high precision described by the following fitting formula:
\begin{gather}
\begin{gathered}
\mathcal{R}^{[1]}_{\rm fit}=\frac{a_1^{5/4}+b_1^{1/2}}{2}\exp\bigl[0.5068-\sqrt{y^2+0.5068^2}\bigr], \\
a_1=1+0.3118 \, y^2,
\qquad
b_1=1+2.556 \, y^2.
\end{gathered}
\end{gather}
Concerning reduction factor $\mathcal{R}^{[3]}$, our calculations indicate that it can be approximated as
\begin{gather}
\begin{gathered}
\mathcal{R}^{(3)}_{\rm fit}=\frac{3\sqrt{3}}{2\pi}e^{-2y}\bigl(a_3-b_3 e^{-y}\bigr), \\
a_3=3 \, y^2 + 2 \, y + 15.68, \\
b_3=2.791 \, y^3 + 7.732 \, y^2+14.06 \, y + 14.47,
\end{gathered}
\end{gather}
where the fitting coefficients are organized so as to accurately reproduce the asymptotic behavior \eqref{R3} of $\mathcal{R}^{[3]}$ at large values of $y$.

Finally, let us estimate the temperatures where proton superconductivity suppresses the strong process \eqref{sp} to the extent that it can no longer be considered equilibrated. To this aim, we compare its reaction rate, $\lambda_{\rm sp}$, with the typical reaction rates $\lambda_{\rm weak}$ of weak nonleptonic processes. In the absence of pairing, the strong process is $14$–$16$ orders of magnitude faster than the weak ones: $\lambda_{\rm sp} \sim (10^{14}\text{--}10^{16}) \lambda_{\rm weak}$ \cite{ofengeimetal2019}. When protons become superconducting, the strong reaction rate is suppressed $\lambda_{\rm sp}\to\mathcal{R}\lambda_{\rm sp}$ by the reduction factor $\mathcal{R}$, which can be estimated as \cite{haenseletal2000}
\begin{gather}
\mathcal{R}\simeq \frac{0.221}{\tau^{3.5}}\exp\biggl(-\frac{1.764}{\tau}\biggr), \quad \tau\equiv\frac{T}{T_{\rm cp}},
\end{gather}
with $T_{\rm cp}$ being the critical temperature of the proton superconductivity onset. Using this expression, it is straightforward to show that for $\tau \lesssim 0.04$ the superconductivity-suppressed strong process becomes even slower than the unsuppressed weak nonleptonic reactions (say, those that do not involve protons). Assuming that the critical temperature takes values in the range $T_{\rm cp} \simeq (1\text{--}5)\times 10^9\, \rm K$, we find that this occurs at local temperatures $T \lesssim (4 \times 10^7 \, \text{--} \, 2\times 10^8 )\, \rm K$. We see that, for neutron stars in LMXBs (see Fig.\,\ref{representative window}), whether the strong process is equilibrated or not may be quite sensitive to the assumed proton superconductivity model. For models with lower critical temperatures $T_{\rm cp}$, the strong process likely remains equilibrated. For those with higher $T_{\rm cp}$, however, this approximation may no longer be valid. 

In reality, whether the strong process is equilibrated or not should have only a minor effect on the $r$-mode instability windows. Indeed, it cannot significantly contribute to the bulk viscosity $\zeta$,  as our estimates show that the corresponding $\zeta_{\rm max}$ is several times smaller than that of the weak nonleptonic reactions. The adiabatic index, in turn, may depend on how the strong process is treated, but its associated variations are likely weak (see, for example, Fig.\,3 of Ref.\,\cite{ghk14} with adiabatic indices assuming either extremely fast or frozen strong processes). In fact, even larger variations of the adiabatic index are not expected to {\it qualitatively} affect the $r$-mode instability windows (see, for example, Sec.\,\ref{Numerical results}, where we investigate how accounting for the effect of weak nonleptonic reactions on the adiabatic index influences instability windows). Finally, we would like to mention that the strong process becomes relevant only in sufficiently massive stars where $\Xi^-$-hyperons are present in the core. Among our models, this condition is satisfied solely by the $1.7\,M_\odot$ star with an FSU2H core (see Sec.\ \ref{Numerical results} for details). All other stellar models considered do not contain $\Xi^-$-hyperons, and thus the issue of treating the strong process does not arise in those cases.


\subsection{Energy dissipation and gravitational radiation}\label{Instability windows}

The equations \eqref{system}, \eqref{Delta epsilon eq}, and \eqref{Delta p equation} offer (at least within approximations made) the most accurate treatment of oscillations in viscous matter. In typical NS conditions, this consideration, however, is excessive: $\eta$ and $\zeta$ affect oscillations at timscales much larger than their period and can therefore be accounted for perturbatively. One of their primary effects is to cause the oscillation energy $E$ to slowly decrease at the rate
\begin{gather}
\dot{E}_{\rm diss}=\dot{E}_\eta+\dot{E}_\zeta,
\end{gather}
where $\dot{E}_\eta$ and $\dot{E}_\zeta$ represent individual contributions of shear and bulk viscosity to the total energy dissipation rate $\dot{E}_{\rm diss}$, respectively. Generally, they can be calculated as the following integrals over the stellar volume (with $d^3x\equiv dr\,d\theta\,d\varphi$) \cite{kgk2024}:
\begin{gather}
\label{EdotZeta}
\dot{E}_\zeta=-\int\zeta(\nabla_\mu \mathfrak{u}^\mu)^2 c^2\frac{\sqrt{-{\rm g}}}{\Lambda} \ d^3x, \\
\label{EdotEta}
\dot{E}_\eta=-\int\frac{\eta}{2}\sigma_{\mu\nu}\sigma^{\mu\nu} c^2 \frac{\sqrt{-{\rm g}}}{\Lambda} \ d^3x,
\end{gather}
where $\rmg\equiv\det\rmg_{\mu\nu}$. Within the perturbative approach, the leading contribution to these energy loss rates is found by using the velocity $\fru^\mu$ (to be more specific, its perturbation $\delta\fru^\mu$ or, equivalently, Lagrangian displacement $\xi^\mu$) determined as solution to the {\it nondissipative oscillation equations}, i.e., those with $\eta$ and $\zeta$ set to zero, but still accounting for the effect of chemical reactions on the adiabatic index $\gamma$,
\begin{gather}
\label{nondissipative equations}
\left\{
\begin{gathered}
\sigma_{\rm loc}^2\xi^\mu-2i c \sigma_{\rm loc}\xi^\rho\nabla_\rho u^\mu=\frac{c^2}{w_0}\biggl[\pnabla^\mu\delta p-\frac{\delta w}{w_0}\pnabla^\mu p_0\biggr] \hfill \\
\Delta\varepsilon+w_0\pnabla_\mu\xi^\mu=0 \hfill \\
\Delta p+\gamma p_0\pnabla_\mu\xi^\mu=0. \hfill
\end{gathered}
\right.\raisetag{1.0cm}
\end{gather}

In this study, in addition to dissipative effects, we are also interested in how the energy $E$ evolves due to emission of gravitational waves that accompanies the NS oscillations. Precise treatment of this effect requires considering linearized Einstein equations instead of hydrodynamic ones, which goes far beyond the scope of this study. Nevertheless, we still can estimate the associated oscillation energy changes, $\dot{E}_{\rm GW}$, by using the standard result of the multipole formalism \cite{thorne1980, chugunov2017}%
\footnote{Note that this formula contains an additional factor $(\sigma_r/\sigma)$. This factor arises because we define the oscillation energy as energy perturbation, measured in the reference frame (co)rotating with angular velocity $\Omega$ (see \cite{chugunov2017}). The original paper \cite{thorne1980}, in turn, considers the reference frame of a distant inertial observer.}:
%
\begin{gather}
\label{EdotGW}
\begin{gathered}
\dot{E}_{\rm GW}=-\frac{\sigma_r}{\sigma}\frac{G}{c^{3}}\sum_{L=2}^\infty \sum_{M=-L}^L \biggl(\frac{\sigma}{c}\biggr)^{2L+2}N_L[|\mathcal{I}_{LM}|^2+|\mathcal{S}_{LM}|^2], \\
N_L=\frac{4\pi}{[(2L+1)!!]^2}\frac{(L+1)(L+2)}{L(L-1)}, \raisetag{1cm}
\end{gathered}
\end{gather}
where $\sigma_r=\sigma+m\Omega$ is the oscillation frequency, measured in the reference frame, rotating with angular velocity $\Omega$, while $\mathcal{I}_{LM}$ and $\mathcal{S}_{LM}$ are the mass and current multipole moments, respectively. Generally, they are given by the integrals of the stress-energy (pseudo)tensor of the {\it entire} physical system, including perturbations of gravitational field (i.e., metric coefficients). In our calculations, we follow a common practice and estimate their values by ignoring perturbations of the gravitational field (in hydrodynamic equations, adopted here, they were anyway discarded as small compared to hydrodynamic perturbations) and using hydrodynamic perturbations, found as solutions to the nondissipative hydrodynamic equations. Here, we do not provide the resulting expressions for the multipole moments and refer the interested reader to original paper \cite{thorne1980} and also to discussion in our recent study \cite{kgk2024}, considering the application of the multipole formalism to relativistic $r$-modes.  We expect the described approximation to provide reasonable energy change rates $\dot{E}_{\rm GW}$ since gravitational radiation, similarly to dissipative effects, weakly affects oscillations and
manifests at timescales much larger than their period.

Formula \eqref{EdotGW} reveals a particularly important feature of oscillations in rotating stars. If a given oscillation has $\sigma_r/\sigma<0$, the emission of gravitational waves results in its amplification ($\dot{E}_{\rm GW}>0$), rather than its damping ($\dot{E}_{\rm GW}<0$). This phenomenon is known in the literature as the Chandrasekhar–Friedman–Schutz (CFS) instability \cite{chandra1970, fs1978_1, fs1978_2, friedman1978}, and it reflects the fact that the emission of gravitational waves drives the transfer of rotational energy of the star into oscillation energy \cite{chugunov2017}. This process competes with the energy dissipation caused by $\eta$ and $\zeta$, and the total oscillation energy change rate is given by the sum
\begin{gather}
\dot{E}=\dot{E}_\eta+\dot{E}_\zeta+\dot{E}_{\rm GW},
\end{gather}
that, depending on stellar parameters, can take both positive and negative values. Generally, $\dot{E}=\dot{E}(\Omega,T^\infty)$ can be viewed as a function of angular velocity $\Omega$ and redshifted stellar temperature $T^\infty$. For certain combinations of these parameters, $(\Omega, T^\infty)$, the amplification of oscillations by the CFS mechanism is exactly balanced by energy dissipation: $\dot{E}(\Omega,T^\infty)=0$. This condition determines the curve on the $(\Omega,T^\infty)$ plane, that divides the plane into two regions: the first, where the energy dissipation prevails over the CFS instability and oscillations are suppressed, and the second, where the CFS mechanism prevails and oscillations become unstable. This second region is known in the literature as the {\it instability window} (see the previously shown Fig.\,\ref{representative window} for a representative example). In what follows, we apply the reviewed theory to investigate the instability windows of relativistic $r$-modes in neutron stars with hyperonic matter in their cores.

\newpage


\section{Application to r-modes}\label{Application to r-modes}

In this section, we apply the previously outlined ideas to calculate relativistic $r$-mode instability windows in hyperonic stars. First, we discuss the peculiar properties of relativistic $r$-modes \eqref{Nonanalytic r-modes} and their influence on the $r$-mode energy change rates \eqref{Energy change rates}. Then we provide the resulting instability windows for various NS models and microphysical conditions \eqref{Numerical results}.


\subsection{Peculiar properties of relativistic $r$-modes}\label{Nonanalytic r-modes}

By definition, $r$-modes belong to the class of oscillations of rotating neutron stars, restored primarily by the Coriolis force. They are often characterized as ``quasitoroidal", meaning that the velocity perturbation $\delta\fru^\mu$ or, equivalently, the Lagrangian displacement $\xi^\mu$ in the case of $r$-modes can be approximately written as (see, e.g., \cite{provost1981, ak2001, low1998})
\begin{gather}
\xi^{\mu}\approx\frac{1}{r\sin\theta}\bigl[\delta^\mu_\theta\p_\varphi-\delta^\mu_\varphi \p_\theta\bigr]\mathrm{T}(r,\theta)e^{i\sigma t+i m\varphi}, \label{toroidalxi}
\end{gather}
To determine the oscillation frequency $\sigma$ and unspecified function $\mathrm{T}(r,\theta)$, we consider the nondissipative oscillation equations \eqref{nondissipative equations}, as discussed in Sec.\,\ref{Instability windows}.

It is customary to expand $\mathrm{T}(r,\theta)$ in associated Legendre polynomials (ALPs) $P_{Lm}(\cos\theta)$. Such a decomposition (in the case of $r$-modes) leads to the separation of variables in the oscillation equations, which allows one to classify the angular dependence of different $r$-modes by the corresponding azimuthal ($m$) and orbital ($l$) quantum numbers. For an $(lm)$-mode, the function $\mathrm{T}(r,\theta)$ and oscillation frequency can be approximately written as
\begin{gather}
\mathrm{T}_{lm}(r,\theta)\approx (-i)T_{lm}(r)P_{lm}(\cos\theta),
\\
\sigma_{lm}\approx\sigma^{(0)}_{lm}\Omega, \quad \sigma^{(0)}_{lm}\equiv\biggl[\frac{2m}{l(l+1)}-m\biggr]\Omega. \label{sigmalm}
\end{gather}
In order to find the {\it toroidal function} $T_{lm}(r)$, it is necessary to account for the nontoroidal components of the motion, specifically the corrections to the approximate formulas \eqref{toroidalxi}--\eqref{sigmalm}. Since the $l=m=2$ mode is expected to be the most CFS-unstable, in what follows we consider only the $l=m$ case and abbreviate $\sigma_{lm}\to \sigma$, $\sigma_{lm}^{(0)}\to\sigma^{(0)}$, and $f_{Lm}\to f_L$, where $f_{Lm}(r)$ are the ALP-expansion coefficients of a function $f(r,\theta)=\sum f_{Lm}(r)P_{Lm}(\cos\theta)$.

{\it In Newtonian theory} (see, e.g., \cite{provost1981}), the Lagrangian displacement and the oscillation frequency divided by $\Omega$, i.e., $\sigma/\Omega$, are typically expressed as Taylor series, with $(\Omega/\Omega_{\rm K})^2$ serving as the expansion parameter.%
\footnote{
For modes with frequency $\sigma \propto \Omega$, the angular velocity $\Omega$ enters the oscillation equations only through even powers, which justifies the choice of $(\Omega/\Omega_{\rm K})^2$ rather than $(\Omega/\Omega_{\rm K})$ as the expansion parameter.
}
The equations \eqref{toroidalxi}--\eqref{sigmalm} actually represent the leading-order terms of these series. Then, the analysis of Newtonian $r$-modes is conducted using {\it perturbation theory}, i.e., by equating the terms in the oscillation equations that have the same powers of $\Omega$. {\it In the barotropic crust}, the equations can be solved analytically, and toroidal function is given by a simple formula
\begin{gather}
\label{Newtonian T}
T_{{\rm barotropic},m}^{\rm Newt}(r)=(r/R)^m.
\end{gather}
{\it In the nonbarotropic core}, the problem ultimately reduces to a system of ordinary differential equations for the toroidal function $T_{m}(r)$ and the coefficient $\xi^r_{m+1}(r)$ in the ALP expansion of $\xi^r$. With appropriate boundary conditions, this system constitutes an eigenvalue problem, where $T_{m}(r)$ and $\xi^r_{m+1}(r)$ serve as eigenfunctions, and the role of the eigenvalue is played by the $r$-mode eigenfrequency correction $\sigma^{(1)}$ defined by the equality
\begin{gather}
\sigma=\Omega[\sigma^{(0)}+\sigma^{(1)}].
\end{gather}
The resulting solutions are conveniently classified by the number of nodes of the toroidal function $T_m(r)$ inside the star. In what follows, we focus primarily on the {\it fundamental} $r$-mode [i.e., the mode with nodeless $T_m(r)$], as it is expected to be the most CFS-unstable. For convenience, below we normalize toroidal function by the condition $T_m(R)=1$.

Note that, in Newtonian theory, the toroidal function $T_{m}(r)$ does not depend on $\Omega$, while $\xi^r_{m+1}\propto\Omega^2$ and $\sigma^{(1)}\propto\Omega^2$ are simply proportional to $\Omega^2$. Another particularly important feature of Newtonian $r$-modes is that nonbarotropicity of the matter rather weakly affects the toroidal function $T_m(r)$ of the {\it fundamental} $r$-mode. As a result, for the fundamental $r$-mode, the function
\begin{gather}
h^{\rm Newt}(r)\equiv r^{-m}T^{\rm Newt}_m\propto T^{\rm Newt}_m/T_{{\rm barotropic},m}^{\rm Newt}
\end{gather}
that ``measures" the influence of nonbarotropicity remains approximately constant throughout the entire star.

\begin{figure*}[t]
\centering
\includegraphics[width=1.0\linewidth]{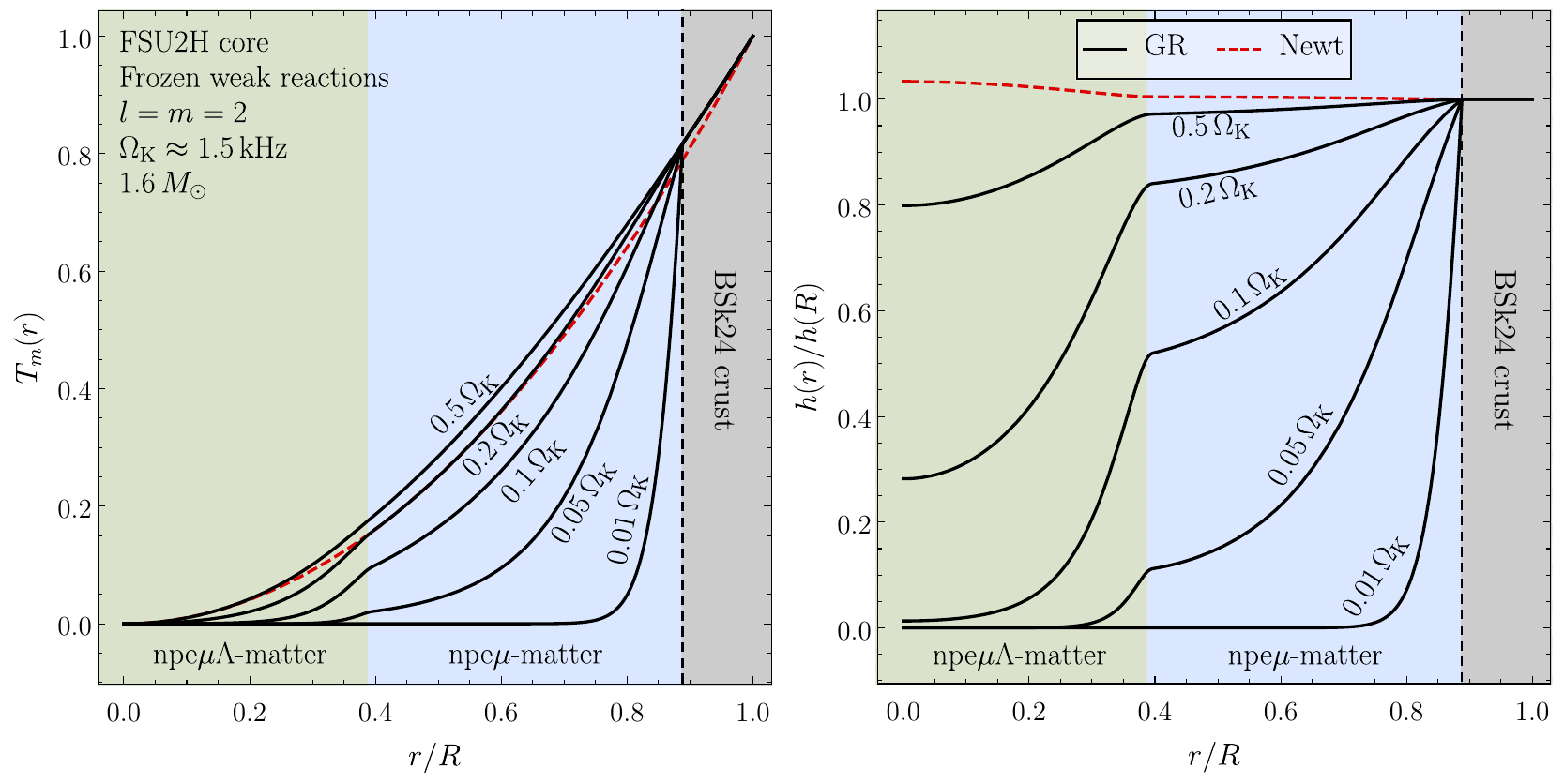}
\vspace{-0.7cm}
\caption{Illustration of relativistic $r$-mode suppression in the neutron star core in the limit of slow rotation. Black curves show the fundamental relativistic $l = m = 2$ $r$-mode toroidal functions $T_m(r)$ [left panel] and the corresponding ratios $h(r)/h(R)$ [right panel], which quantify the effect of matter nonbarotropicity. These functions are computed for angular velocities $\Omega$ ranging from $0.5\,\Omega_{\rm K}$ to $0.01\,\Omega_{\rm K}$. In both panels, red dashed lines represent the corresponding Newtonian solution, which is independent of $\Omega$. Background shading indicates regions with different chemical compositions. The calculation assumes a $1.6\,M_\odot$ neutron star with an FSU2H core and BSk24 crust (see details on microphysical input in Sec.\,\ref{Numerical results}), and is performed in the ``frozen" regime, where weak reactions are negligibly slow and the adiabatic index is given by Eq.\,\eqref{gamma frozen}.}
\label{localization}
\end{figure*}

In contrast to the Newtonian case, the {\it relativistic oscillation equations} require a more delicate approach. The pioneering study of 1997 \cite{kojima1997} revealed that the frame-dragging effect, $\omega(r)$ [see metric tensor \eqref{geometry}], introduces significant complications in modeling relativistic $r$-modes, giving rise to a set of contradictory results now collectively referred to as the {\it ``continuous spectrum problem"} (see \cite{kgk2022_1} for a detailed review). Subsequent attempts to analyze relativistic $r$-modes using perturbation theory have consistently predicted that their frequencies, $\sigma_r = \sigma + m\Omega$, should fill a continuous band.
\begin{gather}
[1-\omega(0)]\sigma_{r,\rm Newt}\leq\sigma_r\leq[1-\omega(R)]\sigma_{r, \rm Newt},
\end{gather}
where $\sigma_{r, \rm Newt}$ are the frequencies of the ``corresponding" Newtonian $r$-modes. Despite that prediction, more general numerical calculations \cite{yl2002, vbh2005}, conducted beyond the perturbation theory, have not found any evidence of a continuous part in the spectrum. Recently,
it has been shown \cite{kgk2022_1, kgk2022_2} that this problem can be resolved by abandoning perturbation theory, thereby allowing for a description of relativistic $r$-modes with nonanalytic functions of $\Omega$. Allowing for such {\it nonanalyticity} complicates the analysis of the equations and requires the use of a number of uncommon techniques. A detailed discussion of these issues lies beyond the scope of the present study; for a comprehensive treatment, the interested reader is referred to \cite{kgk2022_1}. One approximation has to be mentioned though: for simplicity, we treat the frame-dragging function $\omega(r)$ as small, which simplifies the equations and, at the same time, preserves the peculiar features of relativistic $r$-modes (in particular, the continuous spectrum problem holds in this approximation, if perturbation theory is applied). This approximation may not be exactly accurate in the deepest regions of the star, but its accuracy improves as one approaches the NS surface, since, generally, $\omega(r)$ is proven to be a positive monotonically decreasing function of $r$ \cite{hartle1967}.

Within approximations made, modeling relativistic $r$-modes reduces to solving the eigenvalue problem, that resembles Newtonian one and reproduces it in the appropriate limit. The explicit form of the resulting $r$-mode equations is provided in Appendix \ref{Modeling r-modes}. In the crust, which is treated as {\it barotropic} in this study, similarly to Newtonian case, equations can be solved explicitly, and toroidal function is given by
\begin{gather}
\label{GR T}
T_{{\rm barotropic},m}^{\rm GR}(r)=(r/R)^m e^{-(m-1)[\nu(r)-\nu(R)]}.
\end{gather}
This solution represents a straightforward relativistic generalization of the Newtonian result \eqref{Newtonian T}. {\it In the nonbarotropic core}, $r$-modes are described by a system of ordinary differential equations similar to the Newtonian ones. Unlike Newtonian case, however, relativistic equations explicitly contain the angular velocity $\Omega$, that does not vanish from equations if one assumes the Newtonian scalings $\sigma^{(1)}\propto\Omega^2$ and $\xi_{m+1}^r\propto \Omega^2$. The reason is, the eigenfrequency correction $\sigma^{(1)}$ enters equations only through the combination $[\sigma^{(1)}+\sigma_r^{(0)}\omega(r)]T_m(r)$ that, because of the frame-dragging effect $\omega(r)$, does not scale as $\Omega^2$ if one assumes $\sigma^{(1)}\propto\Omega^2$ [see Eq.\,\eqref{system core}]. As a result, despite the similarity of the equations to Newtonian ones, their solutions demonstrate a number of peculiar properties. 

{\it First}, the main contribution to the frequency correction $\sigma^{(1)}$ in GR comes from the frame-dragging effect rather than slow stellar rotation, as it does for Newtonian $r$-modes. {\it Second}, relativistic $r$-mode eigenfunctions nontrivially depend on $\Omega$ and obey a scaling relation $\xi^r_{m+ 1}(r)/T_{m}(r)\sim \Omega$ which does not reduce to simple proportionality $(\propto)$ and significantly deviates from that $\xi^r_{m+1}(r)/T_{m}(r)\propto \Omega^2$ of Newtonian $r$-modes. In particular, as $\Omega$ approaches zero, relativistic $r$-modes exhibit strong suppression in the bulk of the NS core, which was shown \cite{kgk2022_1, kgk2022_2} to be described by an exponential law of the form $e^{f(r)\Omega_{\rm K}/\Omega}$. This asymptotic behavior reveals that, in fact, {\it relativistic $r$-modes are described by nonanalytic functions of $\Omega$} (which is why the perturbation theory fails to describe them). {\it Finally}, unlike Newtonian case, the {\it fundamental} relativistic $r$-mode is strongly affected by the matter nonbarotropicity, and the function
\begin{gather}
h^{\rm GR}(r)\equiv r^{-m}e^{(m-1)\nu(r)}T^{\rm GR}_m\propto T^{\rm GR}_m/T_{{\rm barotropic},m}^{\rm GR}
\end{gather}
significantly varies throughout the star.

We illustrate the described suppression of $r$-modes in the left panel of Fig.\ \ref{localization}, where the toroidal function $T_m(r)$ of the fundamental $l = m = 2$ $r$-mode is shown for angular velocities $\Omega$ ranging from $0.5\,\Omega_{\rm K}$ to $0.01\,\Omega_{\rm K}$. The right panel of Fig.\,\ref{localization} shows the ratios $h^{\rm GR}(r)/h^{\rm GR}(R)$ [solid black lines, ``GR"] and $h^{\rm Newt}(r)/h^{\rm Newt}(R)$ [red dashes, ``Newt"], that demonstrate the influence of the matter nonbarotropicity on relativistic and Newtonian $r$-modes, respectively. The figure shows particular results for a $1.6\, M_\odot$ star with FSU2H core and BSk24 crust, obtained assuming that weak processes are frozen during oscillations. Changing EOS, mass of the star or accounting for the effect of chemical reactions on the adiabatic index does not influence the {\it qualitative} behavior of $r$-modes. However, these changes do affect $r$-mode eigenfunctions {\it quantitatively}, which influences their energy change rates and, therefore, instability windows, as we demonstrate below.


\subsection{Relativistic r-mode energy change rates}\label{Energy change rates}

We remind the reader that modeling instability windows requires comparing the efficiency of the CFS-mechanism $\dot{E}_{\rm GW}$, which supplies energy to the $r$-modes, with the $r$-mode energy losses $\dot{E}_{\rm diss}=\dot{E}_{\eta}+\dot{E}_\zeta$ due to shear viscosity and bulk viscosity. Explicit expressions for the relevant energy change rates can be obtained by following the approach outlined in Sec.\ \ref{Instability windows}, as was done in our previous study \cite{kgk2024}. Specifically, we have shown that at slow rotation rates the leading contributions to the $r$-mode energy change rates take the following form:
\begin{widetext}
\begin{gather}
\label{All Edot}
\begin{gathered}
\dot{E}_{\rm GW}(\Omega,T^\infty)=32\pi G \alpha^2 \frac{2^{2 m} (m-1)^{2 m}}{(m+1)^4}\left[\frac{(m+2)!}{(2 m+1)!}\right]^2\left(\frac{m+2}{m+1}\right)^{2 m}\frac{\Omega ^2}{c}\left(\frac{\Omega }{c}\right)^{2 m+2}\biggl[\int\limits_0^R  \frac{w_0(r)}{c^2} T_m(r,\Omega,T^\infty) e^{2 \lambda(r)} r^{m+2} dr \biggr]^2, \hfill \\
\dot{E}_\eta(\Omega,T^\infty)=-\alpha^2\Omega ^2\int\limits_0^{r_{\rm cc}} \eta(r,T^\infty) \bigl\{\left[T_m(r,\Omega,T^\infty)-r T_m'(r,\Omega,T^\infty)\right]^2+(m-1) (m+2) e^{2 \lambda(r)} T_m^2(r,\Omega,T^\infty)\bigr\}e^{-\lambda(r)} dr, \hfill \\
\dot{E}_\zeta(\Omega,T^\infty)=-\frac{16 m \alpha^2 \Omega ^6}{c^4 (m+1)^5 (2 m+3)}\int\limits_0^{r_{\Lambda}}  \frac{\zeta(r,\Omega,T^\infty)}{A^2(r,\Omega,T^\infty)}\left[\frac{w_0(r)h'(r,\Omega,T^\infty)}{p_0(r)\gamma(r,\Omega,T^\infty)}\right]^2 e^{\lambda(r)-2(m+1) \nu(r)} r^{2 m+4} \ dr. \hfill
\end{gathered}
\raisetag{0.9cm}
\end{gather}
\end{widetext}
Here, we abbreviated $h^{\rm GR}$ as $h$, the prime $'$ denotes the derivative $\p/\p r$, $\alpha$ is some constant, proportional to the $r$-mode amplitude, and $A(r,\Omega,T^\infty)$ is the so-called Schwarzschild discriminant, defined as
\begin{gather}
\label{A}
A(r,\Omega,T^\infty)=\frac{w_0'(r)}{w_0(r)}-\frac{p_0'(r)}{w_0(r)}\biggl[1+\frac{w_0(r)/p_0(r)}{\gamma(r,\Omega,T^\infty)}\biggr].
\end{gather}
In formulas for $\dot{E}_{\eta,\zeta}$ we ignore the contribution of the NS crust. In the case of $\dot{E}_\eta$, we integrate over the entire NS core: $0\leq r\leq r_{\rm cc}$, where $r=r_{\rm cc}$ corresponds to the crust-core interface. In the case of $\dot{E}_\zeta$, since we consider only hyperonic bulk viscosity, the integration is restricted to the region of the core containing hyperons: $0 \leq r \leq r_\Lambda$, where $r_\Lambda$ is the radius at which $\Lambda$-hyperons first appear (i.e., the point where the composition changes from $\rm npe\mu$ to $\rm npe\mu\Lambda$). Ignoring the crust is justified since dissipation therein is provided mainly by shear viscosity, which is substantially less efficient than hyperonic bulk viscosity in the NS core. 

Note that we explicitly include the dependence on $\Omega$ and $T^\infty$ in the arguments of all relevant functions, participating in expressions \eqref{All Edot}. By writing that shear viscosity $\eta$, bulk viscosity $\zeta$, and adiabatic index $\gamma$ depend on $T^\infty$ and $\Omega$ we imply that these quantities actually depend on the local temperature $T(r)$ of the matter and local oscillation frequency $\sigma_{\rm loc}(r,\Omega)$, that should be related to the constant redshifted temperature $T^\infty$ and oscillation frequency $\sigma$ by formulas \eqref{thermal equilibrium} and \eqref{Lagrangian displacement}. For simplicity, we calculate $\zeta_{\rm max}$ and $\lambda_{\rm max}$ using the approximate local $r$-mode frequency $\sigma_{\rm loc}(r,\Omega)\approx\Lambda(r)[\sigma^{(0)}(\Omega)+m\Omega]$ with $\sigma^{(0)}$ given by \eqref{sigmalm}. Next, the temperature dependence of the toroidal function $T_m(r,\Omega,T^\infty)$ and function $h(r,\Omega,T^\infty)$ arises because $r$-mode equations involve adiabatic index $\gamma(r,\Omega,T^\infty)$. However, in the limiting cases of frozen matter or extremely fast reactions, discussed at the end of Sec.\,\ref{Bulk viscosity and adiabatic index}, adiabatic index becomes temperature-independent, and so does the toroidal function. Finally, the dependence of $T_m(r, \Omega, T^\infty)$ and $h(r, \Omega, T^\infty)$ on $\Omega$ arises both from the nonanalytic nature of relativistic $r$-modes, as discussed in Sec.\ \ref{Nonanalytic r-modes}, and from the $\Omega$-dependence of the adiabatic index $\gamma(r, \Omega, T^\infty)$, which enters the $r$-mode equations.

The expressions \eqref{All Edot} for the $r$-mode energy change rates provide a straightforward relativistic generalization of their Newtonian counterparts and reduce to them in the appropriate limit (which involves replacing $h=h^{\rm GR}\to h^{\rm Newt}$). Despite this similarity, the peculiar properties of relativistic $r$-modes may lead to a rather complicated dependence of their energy change rates on angular velocity $\Omega$, compared to the Newtonian case.

{\it First}, in the limit of slow rotation, all rates are modified primarily by suppression of $r$-modes in the NS core. Because of the suppression, the core contributes to $\dot{E}_{\eta,\rm GW}$ only through a narrow region near the crust-core interface $r_{\rm cc}$, while the dissipation $\dot{E}_\zeta$ through bulk viscosity, taking place in a deeper region $0\leq r\leq r_{\Lambda}$, becomes negligibly small. The width of the contributing region near $r_{\rm cc}$ depends itself on $\Omega$ and, therefore, modifies the $\Omega$-dependence of the energy change rates in the core \cite{kgk2024}. Moreover, in this region, $r$-modes exhibit steep exponential $e^{f(r)\Omega_{\rm K}/\Omega}$ growth (see Fig.\,\ref{localization}), which leads to extremely large derivatives: $T_m'\sim (\Omega_{\rm K}/\Omega)T_m$ and $h'\sim (\Omega_{\rm K}/\Omega)h$. This effect drastically influences the energy dissipation rate $\dot{E}_{\eta}$, which effectively ``acquires" an additional $(\Omega_{\rm K}/\Omega)^2$ factor. Note, however, that rotation rates where $r$-mode suppression takes place are typically lower than those observed in LMXBs and, therefore, their influence at relevant angular velocities is not that strong.

{\it Second}, as we mentioned in the previous section, relativistic $r$-modes are modified by the matter nonbarotropicity much stronger than the Newtonian ones. We demonstrated this effect by the right panel of Fig.\,\ref{localization} where Newtonian function $h^{\rm Newt}$, that ``measures'' the effect of nonbarotropicity, remains approximately constant, while its relativistic counterpart $h^{\rm GR}$ noticeably varies throughout the stellar core. Unlike $r$-mode suppression, taking place only at extremely slow rotation rates, this effect operates also at relatively high angular velocities $0.1\,\Omega_{\rm K}\lesssim\Omega\lesssim 0.5 \, \Omega_{\rm K}$, where relativistic $r$-modes are not suppressed in the NS core and, from this viewpoint, resemble Newtonian ones. At these rotation rates, we find that in the region $0\leq r \leq r_\Lambda$ the ratio $|h^{\rm GR}{}'/h^{\rm Newt}{}'|^2$ may reach values as high as $\sim 3 \times 10^3$, indicating that, according to \eqref{All Edot}, relativistic dissipation through bulk viscosity is immensely amplified compared to Newtonian one. This effect is of paramount importance for the $r$-mode instability windows, since, unlike the mode suppression, it operates at rotation rates typical for LMXBs. This effect also operates at lower rotation rates, but the $r$-mode suppression therein prevails and the dissipation through bulk viscosity still becomes negligible compared to that through shear viscosity.

\vfill


\nopagebreak

\subsection{Instability windows}\label{Numerical results}


\subsubsection{Microphysical input}

In our calculations, we model the NS crust using BSk24 EOS \cite{gorielyetal2013}. Specifically, we employ fits \cite{pearsonetal2018} for the pressure $p_{\rm BSk24}(\rho)$ as a function of density $\rho=\varepsilon/c^2$, corresponding to the BSk24 crust in chemical equilibrium. To describe the core, we use either FSU2H EOS from \cite{providenciaetal2019} or TM1C EOS from \cite{ghk14}, both based on relativistic mean-field theory. We define the crust-core interface as corresponding to the density $\rho_{\rm cc}$, where the equilibrium BSk24 pressure matches the equilibrium pressure of the FSU2H or TM1C EOS: $p_{\rm BSk24}(\rho_{\rm cc})=p_{\rm FSU2H/TM1C}(\rho _{\rm cc})$. We expect the FSU2H EOS to yield the most optimistic results for stabilizing $r$-modes in LMXBs, since hyperons in this model appear at relatively low matter densities $\rho$, and even the canonical $M=1.4\, M_\odot$ neutron star already contains a tiny fraction of $\Lambda$-hyperons in its core. In contrast, the TM1C EOS likely represents a more pessimistic scenario, since, in this case, hyperons emerge at higher densities reached in significantly more massive stars with $M\geq 1.55\, M_\odot$. This is particularly true in Newtonian $r$-mode theory, where stabilizing $r$-modes in TM1C model requires substantially higher stellar masses compared to the FSU2H model \cite{ofengeimetal2019}. Here, for the FSU2H EOS in the core, we considered NS models with masses $M = 1.5\,M_\odot$, $1.6\,M_\odot$, and $1.7\,M_\odot$, while for the TM1C EOS, we chose to consider more massive stars with $M = 1.7\,M_\odot$ and $1.8\,M_\odot$. Of all considered models, only the $1.7\,M_\odot$ FSU2H model contains both $\Lambda$- and $\Xi^-$-hyperons in the core, whereas the cores of the remaining models contain only $\Lambda$-hyperons.

\begin{figure}
\centering
\includegraphics[width=1.0\linewidth]{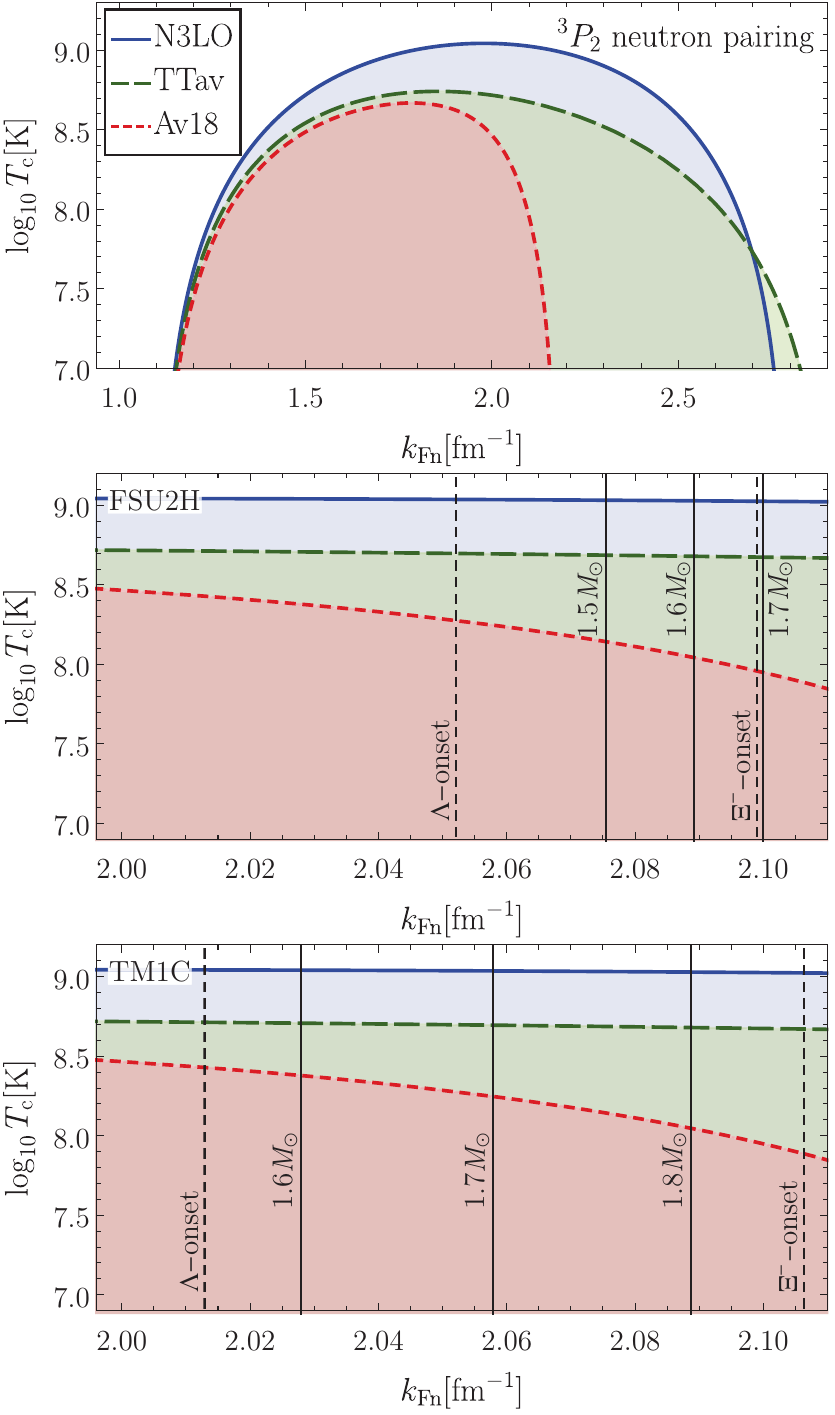}
\vspace{-0.6cm}
\caption{Critical temperatures $T_{\rm c}$ for the onset of neutron superfluidity via the triplet-state ${}^3P_2$ pairing channel, shown as functions of the neutron Fermi wavevector $k_{{\rm Fn}}$. Different colors correspond to different neutron superfluidity models. Shaded regions below each curve correspond to neutrons being in superfluid state. The first panel shows $T_{\rm c}(k_{\rm Fn})$ over a broad range of Fermi wavevectors. The second and third panels provide zoomed-in views highlighting Fermi momenta corresponding to the onset of hyperons (black dashed lines) and to the central densities of the stellar models considered (black solid lines). Specifically, the second panel corresponds to the FSU2H EOS, while the third corresponds to the TM1C EOS.}
\label{critical_temperatures_fig}
\end{figure}

To calculate bulk viscosity $\zeta$ and adiabatic index $\gamma$, we use the fits for the reaction rates $\lambda_{12\leftrightarrow 34}$ of weak nonleptonic reactions from the work \cite{ofengeimetal2019} that accounts for the most efficient channel for these reactions, namely, meson exchange. Next, concerning shear viscosity $\eta$, we estimate its values by adopting the relevant fits for normal and strongly superconducting nucleonic matter from \cite{ss2018, dommes2020}. Finally, regarding neutron superfluidity, we employ three different neutron pairing models. First, we use the fit from \cite{Hoetal2015} for the energy gap of the TTav neutron pairing model developed in \cite{tt2004}. Second, we use the fits from \cite{drddwcp16} for the pairing gaps calculated assuming the Av18 \cite{wss1995} and N3LO \cite{em2003} nucleon interaction models.%
\footnote{
Note that this approach is not self-consistent, as the EOS and pairing gaps, ideally, should be calculated assuming the same nucleon interaction potential.
}.
The corresponding neutron critical temperatures $T_{\rm c}(k_{\rm Fn})$ as functions of the neutron Fermi wave vector $k_{\rm Fn}$ are shown in the first panel of Fig.\,\ref{critical_temperatures_fig}. The remaining panels of the figure zoom in to show the wave vectors corresponding to the onset of hyperons (black dashed vertical lines) and to the central densities of the stellar models considered (black solid vertical lines). The second panel corresponds to the FSU2H EOS, while the third assumes TM1C EOS.

To obtain the instability windows, we, following the procedure from Appendix \ref{Modeling r-modes}, for each of the considered NS models calculated relativistic $r$-modes and their energy change rates $\dot{E}_{{\rm GW},\eta,\zeta}(\Omega,T^\infty)$, covering a wide range of stellar rotation rates $0.01\,\Omega_{\rm K}\leq \Omega \leq 0.6 \, \Omega_{\rm K}$ and temperatures $10^{6.5}\,{\rm K}\leq T^\infty\leq 10^9\,{\rm K}$. The calculations were performed assuming different combinations of microphysical conditions, which we divided in the following groups:
\begin{itemize}
\item[(1)]
{\it the state of protons}, which may be either normal (``N") or strongly superconducting (``SSc");

\item[(2)]
{\it the state of neutrons}, which may be either normal (``N") or superfluid within models N3LO, Av18, or TTav (``SfN3LO",``SfAv18", or``SfTTav");

\item[(3)]
{\it the way weak nonleptonic reactions are treated} in calculations of the adiabatic index, which may

\begin{itemize}
\item
fully account for their effect (``exact") and use adiabatic index \eqref{gamma};

\item
partially account for chemical reactions assuming that they are extremely fast (``fast") and use adiabatic index \eqref{gamma fast};

\item
completely ignore the effect of chemical reactions (``frozen") and use adiabatic index \eqref{gamma frozen}.

\end{itemize}

For brevity, below we refer to these calculations and resulting $r$-modes simply as ``exact", ``fast", or ``frozen".
\end{itemize}
%


\subsubsection{Numerical results}
\begin{figure*}[t]
\centering
\includegraphics[width=1.0\linewidth]{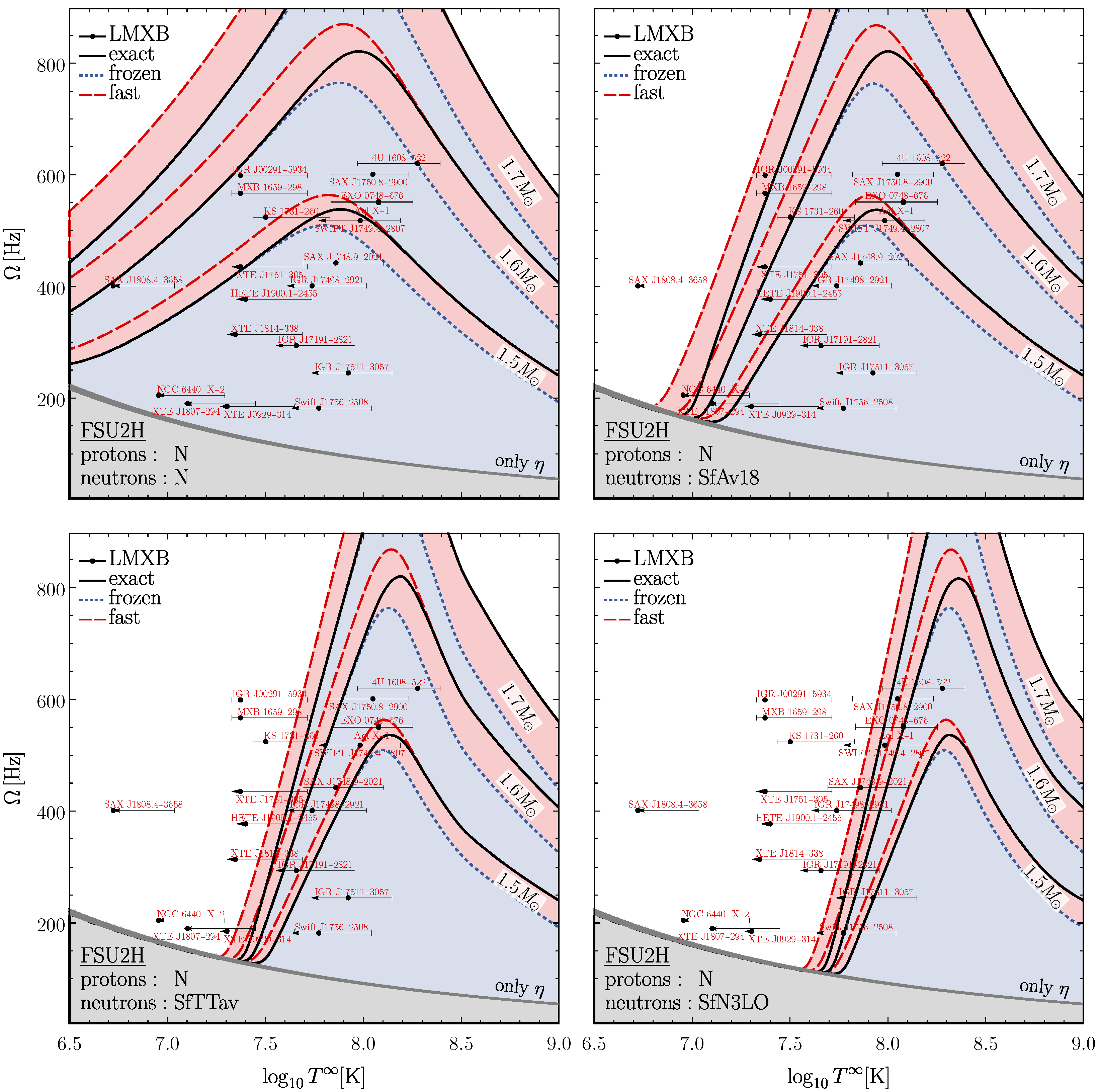}
\vspace{-0.6cm}
\caption{Relativistic $r$-mode instability windows calculated using the FSU2H EOS for the NS core. In each panel, protons are assumed to be normal, while neutrons are considered either normal (upper left panel) or superfluid within models Av18 (upper right panel), TTAv (lower left panel), and N3LO (lower right panel). The windows are presented for stellar masses $M=1.5\,M_\odot$, $M=1.6\, M_\odot$, and $M=1.7\, M_\odot$. For each mass, blue dots (``frozen"), red dashes (``fast"), and solid black line (``exact") show the instability window boundaries for ``frozen", ``fast", and ``exact" $r$-modes, respectively. Gray curve (``only $\eta$"), in turn, shows the boundary of the instability window for ``frozen" $r$-modes with bulk viscosity being ``switched off" and shear viscosity being the only dissipative mechanism. Stability regions of ``frozen" $r$-modes are shaded in blue, while red-shaded bands show their broadening in the case when chemical reactions are treated as extremely fast (i.e., in the case of ``fast" $r$-modes). Gray regions show the stability regions of ``frozen" $r$-modes calculated with shear viscosity being the only dissipative mechanism.  These regions are weakly affected by stellar mass and, hence, are virtually indiscernible. Finally, black dots show the observational data on neutron stars in LMXBs (taken from \cite{kgd2021}), and error bars display the uncertainty arising from the chemical composition of the outer NS layers (see details in \cite{gck2014}).}
\label{FSU2Hnormal}
\end{figure*}

\begin{figure*}[t]
\centering
\includegraphics[width=1.0\linewidth]{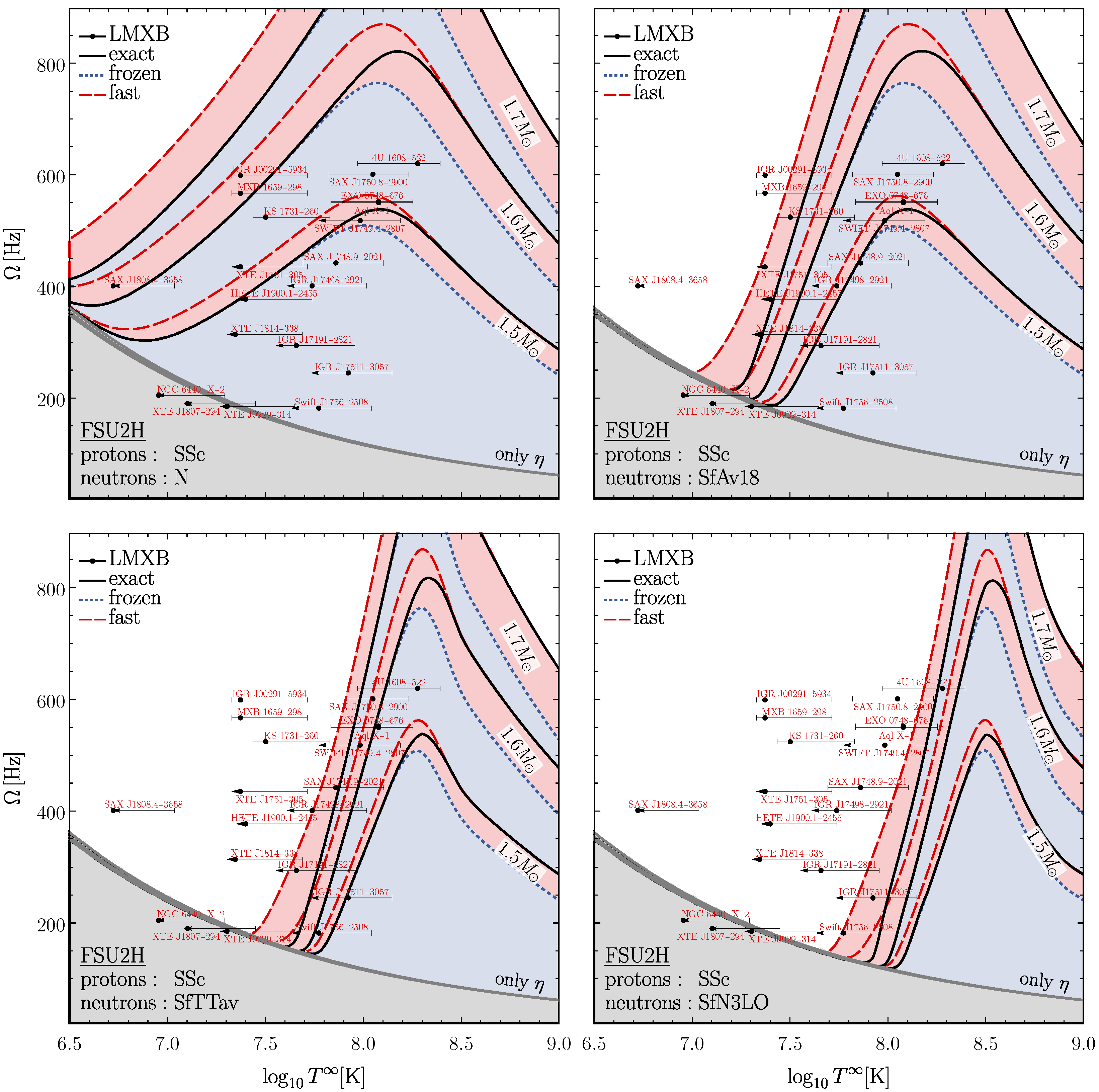}
\vspace{-0.6cm}
\caption{Same as Fig.\,\ref{FSU2Hnormal} but for the case of strongly superconducting protons.}
\label{FSU2Hsupercond}
\end{figure*}

\begin{figure*}[t]
\centering
\includegraphics[width=1.0\linewidth]{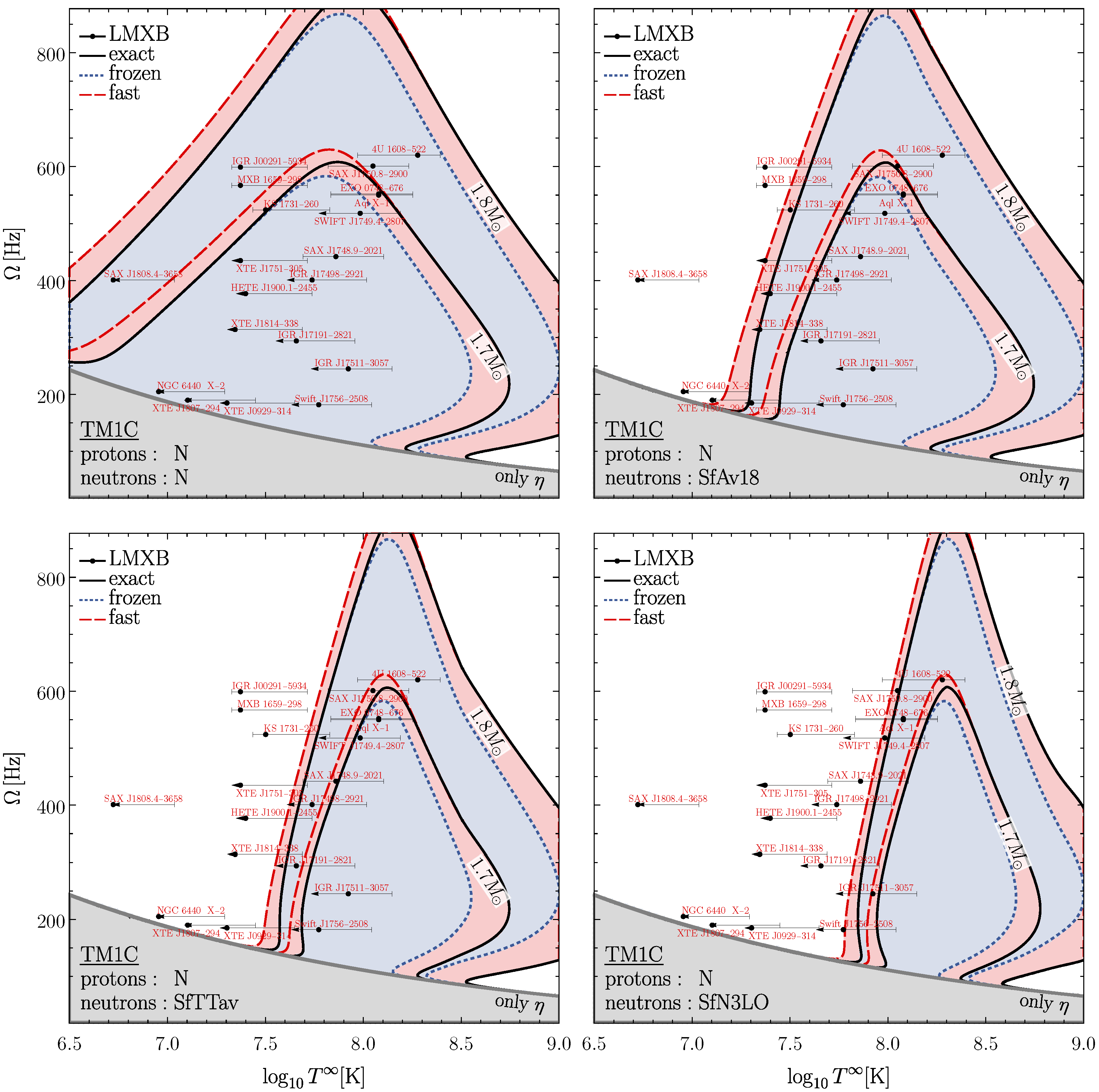}
\vspace{-0.6cm}
\caption{Same as Fig.\,\ref{FSU2Hnormal} but for the neutron star with TM1C core and masses $M=1.7\, M_\odot$ and $M=1.8\, M_\odot$.}
\label{TM1Cnormal}
\end{figure*}

\begin{figure*}[t]
\centering
\includegraphics[width=1.0\linewidth]{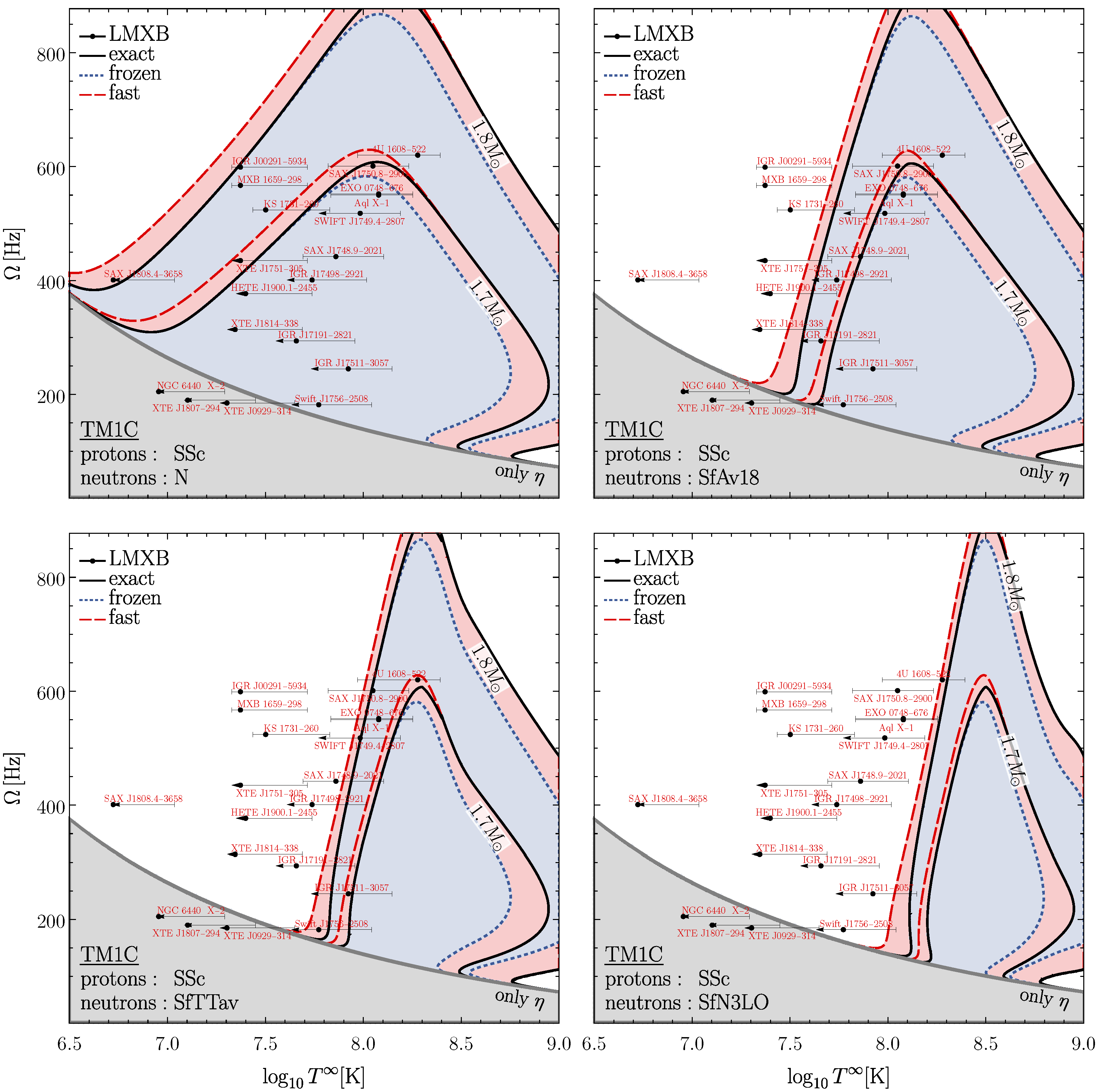}
\vspace{-0.6cm}
\caption{Same as Fig.\,\ref{TM1Cnormal} but for the case of strongly superconducting protons.}
\label{TM1Csupercond}
\end{figure*}

We present the obtained $r$-mode instability windows in Figures \ref{FSU2Hnormal}--\ref{TM1Csupercond}, each corresponding to a specific core EOS and proton state. Figures \ref{FSU2Hnormal} and \ref{FSU2Hsupercond} show results for the FSU2H EOS with normal and strongly superconducting protons, respectively. Likewise, Figures \ref{TM1Cnormal} and \ref{TM1Csupercond} display results for the TM1C EOS for the same proton states. Each figure consists of four panels, corresponding to four different neutron superfluidity models: normal neutrons (upper left), superfluid Av18 neutrons (upper right), superfluid TTav neutrons (lower left), and superfluid N3LO neutrons (lower right). Each panel, in turn, shows the instability windows for neutron stars of different masses. For the FSU2H EOS, the masses considered are $1.5\,M_\odot$, $1.6\,M_\odot$, and $1.7\,M_\odot$. For the TM1C EOS, we choose $1.7\,M_\odot$ and $1.8\,M_\odot$ stellar models.

To visualize the instability windows, for each stellar model we plot four different curves, defined as follows:
\begin{gather*}
\begin{gathered}
\text{\underline{(1) blue dots (``frozen"):}} \hfill \\
\text{\underline{(2) red dashes (``fast"):}} \hfill \\
\text{\underline{(3) black curve (``exact"):}} \hfill  \\
\text{\underline{(4) grey curve (``only $\eta$"):}} \hfill 
\end{gathered}
\quad
\begin{gathered}
\dot{E}_{\rm frozen}(\Omega,T^\infty)=0  \hfill \\
\dot{E}_{\rm fast}(\Omega,T^\infty)=0   \hfill \\
\dot{E}_{\rm exact}(\Omega,T^\infty)=0  \hfill \\
\dot{E}_{\rm frozen}(\Omega,T^\infty)|_{\zeta=0}=0.  \hfill
\end{gathered}
\end{gather*}
Here, the labels ``frozen", ``fast", and ``exact" refer to the corresponding values of the adiabatic index assumed in the calculations. By definition, the first three curves show the boundaries of the instability windows of ``frozen", ``fast" and ``exact" $r$-modes. The fourth curve illustrates the role of shear viscosity in damping $r$-modes and shows the boundary of the instability window, where shear viscosity is assumed to be the sole dissipative mechanism and the effects of weak reactions are completely neglected (i.e., bulk viscosity is ``switched off" and adiabatic index is taken to be ``frozen"). Black dots represent observations of neutron stars in LMXBs taken from \cite{kgd2021}, while the corresponding error bars reflect theoretical uncertainties associated with the chemical composition of the outer NS layers \cite{gck2014}.

We use color shading to distinguish $r$-mode instability windows $(\dot{E}>0)$ from their stability regions $(\dot{E}<0)$. First, the stability regions of ``frozen" $r$-modes are shown in blue,
\begin{gather}
\text{\underline{blue regions:}}
\qquad
\dot{E}_{\rm frozen}(\Omega,T^\infty)\leq 0.
\end{gather}
Next, we find that accounting for the effect of chemical reactions on the adiabatic index, generally, leads to broader stability regions compared to the ``frozen" ones. This broadening is most prominent in the limit of extremely fast reactions -- corresponding to ``fast" $r$-modes -- and is highlighted by red-shaded bands,
\begin{gather}
\text{\underline{red bands:}}
\qquad
\left\{
\begin{gathered}
\dot{E}_{\rm fast}(\Omega,T^\infty) \leq 0 \hfill \\
\dot{E}_{\rm frozen}(\Omega,T^\infty)\geq 0. \hfill
\end{gathered}
\right.
\end{gather}
As follows from the definition, this region shows stellar parameters where the ``frozen" $r$-modes are CFS-unstable while the ``fast" $r$-modes, nevertheless, remain stable. Finally, the regions where $r$-modes are stabilized solely by shear viscosity are shown in gray:
\begin{gather}
\text{\underline{gray regions:}} \qquad \dot{E}_{\rm frozen}(\Omega,T^\infty)|_{\zeta=0}\leq 0.
\end{gather}
These gray regions and their boundaries are weakly affected by stellar mass and, hence, are virtually indiscernible.

\subsubsection{Discussion}

First of all, our calculations confirm that the FSU2H EOS is more favorable than the TM1C EOS for stabilizing $r$-modes in LMXBs. This outcome is expected, as the FSU2H EOS allows hyperons to appear at significantly lower matter densities than the TM1C EOS. In other words, a given FSU2H model contains more hyperons than TM1C model of the same stellar mass. As a result, explaining $r$-mode stabilization in LMXBs in the FSU2H case requires lower neutron star masses than in the TM1C case. From this viewpoint, our results resemble Newtonian ones \cite{ofengeimetal2019}. In our case, however, relativistic effects substantially amplify dissipation through bulk viscosity (recall discussion in Sec.\,\ref{Energy change rates}), leading to $r$-mode stabilization at significantly lower stellar masses compared to those required in the Newtonian case. For example, within Newtonian FSU2H models, in the absence of pairing effects, stabilization of $r$-modes in LMXBs requires masses $M\gtrsim 1.7\, M_\odot$, while proton and especially neutron pairing likely lead to even higher required values $M\gtrsim 1.8 M_\odot$ (see left column of Fig.~12 in \cite{ofengeimetal2019}). Relativistic consideration we provide, in turn, allows stabilization at $M\gtrsim 1.6\,M_\odot$, possibly even when protons are strongly superconducting and neutrons are superfluid (see our FSU2H instability windows with Av18 neutrons). Concerning our TM1C models, $r$-mode stabilization without pairing effects requires $M \gtrsim 1.7\, M_\odot$, while the inclusion of neutron pairing pushes this threshold {\it at least} up to approximately $M \gtrsim 1.8\, M_\odot$. For reference, stabilization in Newtonian TM1C models without pairing requires $M\gtrsim 1.8\, M_\odot$, while neutron and proton pairing increase this threshold up to approximately $M\gtrsim 1.9\,M_\odot$ (see right column of Fig.~12 in \cite{ofengeimetal2019}). We, therefore, conclude that, compared to Newtonian case, hyperonic bulk viscosity in GR is a way more effective dissipative mechanism, capable (depending on the microphysical input) of stabilizing $r$-modes in LMXBs even in the presence of nucleon pairing effects.

Second, our calculations indicate that, independently of the chosen core EOS, proton superconductivity has only a minor impact on the instability windows: it slightly narrows the stability regions and shifts them towards slightly higher temperatures. The weak influence of proton superconductivity is expected, as it suppresses only two [\eqref{npLp} and \eqref{wnpLast}] of the relevant chemical reactions, while the reactions left [\eqref{wnpFirst}, \eqref{nLLL}, and \eqref{nXLX}] remain active. In contrast, neutron superfluidity has a strong effect on the instability windows, as neutrons participate in all considered reactions. Locally, neutron superfluidity severely inhibits reactions as the temperature $T(r)$ in a given point $r$ falls below the corresponding critical temperature $T_{\rm c}(r)$. Globally, this results into a steep left-hand boundary of the $r$-mode stability regions. The exact temperatures $T^\infty$, corresponding to this boundary, depend on the assumed model of neutron superfluidity, specifically on the critical temperature profile $T_{\rm c}(r)$ in the hyperonic core. Physically, as the temperature $T^\infty$ decreases, the core develops a superfluid region with $T(r)< T_{\rm c}(r)$, and this region grows as $T^\infty$ continues to fall. Chemical reactions in this region gradually ``freeze" with decreasing $T^\infty$, which leads to weaker dissipation through bulk viscosity. Eventually, the entire core becomes superfluid, weak reactions become frozen, and dissipation through $\zeta$ becomes negligibly small. Thus, larger critical temperatures $T_{\rm c}(r)$ in the hyperonic core lead to larger redshifted temperatures $T^\infty$, corresponding to the discussed boundary. Among the models considered, the Av18 model has the lowest critical temperatures $T_{\rm c}(r)$ in the hyperonic core (see lower two panels in Fig.\,\ref{critical_temperatures_fig}), making it the most favorable for stabilizing $r$-modes in LMXBs. Remaining TTav and especially N3LO models, in turn, with their higher critical temperatures, struggle to stabilize $r$-modes in LMXBs.

Third, as we already mentioned,  accounting for the effect of chemical reactions on the adiabatic index broadens the $r$-mode stability regions. To explain this effect, we note that chemical reactions tend to reduce the adiabatic index. According to relation $\Delta p=(\gamma p_0/w_0)\Delta\varepsilon$, that follows from \eqref{nondissipative equations}, this implies that they soften the matter: lower pressure $\Delta p$ is required to achieve the same compression $\Delta\varepsilon$. Since the matter becomes easier to compress/decompress, and the latter leads to dissipation through bulk viscosity, this explains why accounting for chemical reactions in adiabatic index leads to stronger energy dissipation and, hence, broader stability regions. Red-shaded bands highlight the resulting difference between stability regions corresponding to limiting cases of ``fast" and ``frozen" $r$-modes. The actual (``exact") boundaries of the stability regions (black curves), which {\it accurately} incorporate the effect of chemical reactions on the adiabatic index, are located between the boundaries of the ``frozen" (blue dashes) and ``fast" (red dashes) stability regions. Specifically, at low temperatures, where chemical reactions are slow, the ``exact" boundaries coincide with the ``frozen" ones. At high temperatures, where reactions are fast, they naturally coincide with the ``fast" boundaries. Finally, in comparatively narrow regions of intermediate temperatures, where reactions proceed at moderate rates, the ``exact" boundaries occupy the intermediate locations. Note that dissipation is most efficient at these intermediate temperatures (recall that bulk viscosity peaks at moderate reaction rates and vanishes if reaction rates are either too slow or too fast, as discussed in Sec.\,\ref{Bulk viscosity and adiabatic index}), which results in a bell-like shape of stability regions. Our results demonstrate that changes in the adiabatic index caused by weak reactions cannot qualitatively affect the instability windows and lead to a relative shift $\lesssim 10\%$ in location of their boundaries.

\begin{figure}
\centering
\includegraphics[width=1.0\linewidth]{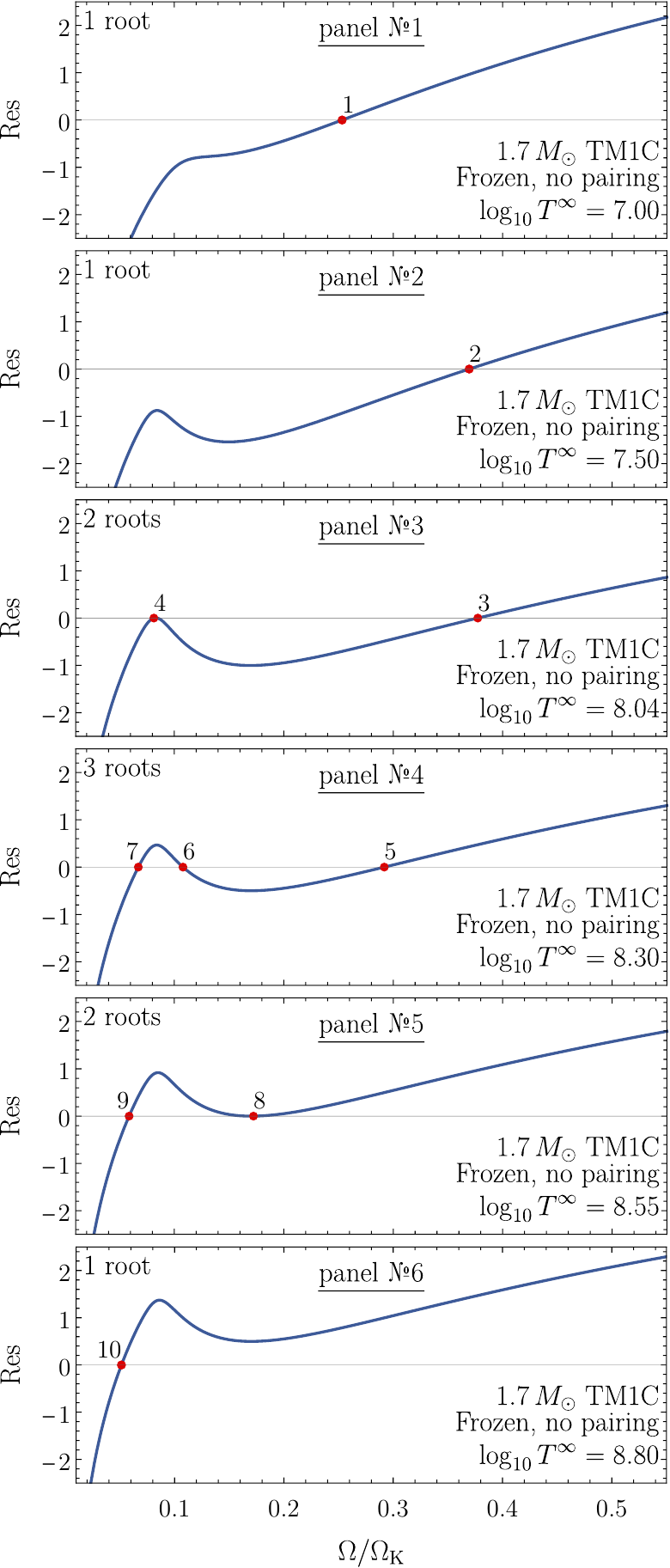}
\vspace{-0.6cm}
\caption{Residual function ${\rm Res}(\Omega,T^\infty)$ as a function of $\Omega$ at different temperatures $T^\infty$ for frozen $r$-modes in the $1.7 \,M_\odot$ TM1C model. Protons and neutrons are taken to be normal. In each panel, red dots show the roots of ${\rm Res}(\Omega,T^\infty)$ for the corresponding temperatures. Panels 1, 2, and 6 show temperatures where ${\rm Res}(\Omega,T^\infty)$ has only one root, panels 3 and 5 illustrate the cases with two roots, and panel 4 shows the case with three roots.}
\label{roots}
\end{figure}

\begin{figure}
\centering
\includegraphics[width=1.0\linewidth]{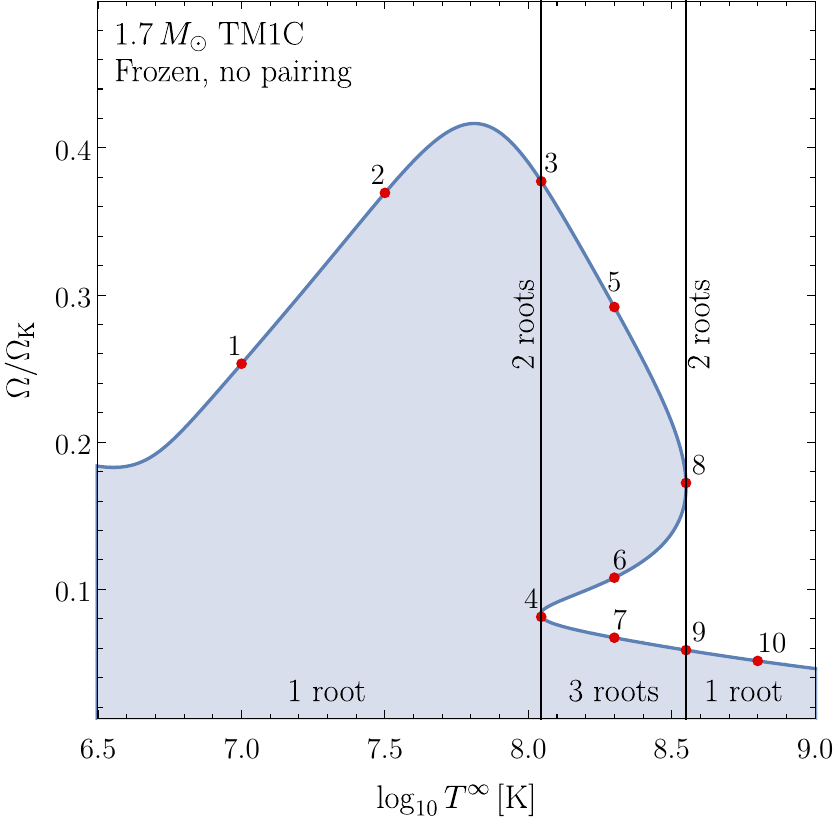}
\vspace{-0.6cm}
\caption{Instability window of ``frozen" $r$-modes in the $1.7\,M_\odot$ TM1C model with normal protons and neutrons. Red dots correspond to the roots of the residual function ${\rm Res}(\Omega,T^\infty)$, shown in Fig.\,\ref{roots}. The numeration of the roots is preserved. Vertical lines, where ${\rm Res}(\Omega,T^\infty)$ has two roots, separate the regions, where ${\rm Res}(\Omega,T^\infty)$ has either one or three roots. }
\label{windowRoots}
\end{figure}

Finally, we would like to pay attention to the nontrivial shape of the instability windows in TM1C models. This is a purely relativistic feature that is absent in Newtonian theory. It emerges due to nontrivial $\Omega$-dependence of the relativistic $r$-mode eigenfunctions that leads to nontrivial $\Omega$-dependence of the energy change rates. This dependence is illustrated in Fig.\,\ref{roots}, where we show the difference
\begin{gather}
\label{Res}
{\rm Res}(\Omega,T^\infty)\equiv\log_{10}\dot{E}_{\rm GW}-\log_{10}|\dot{E}_{\rm diss}|
\end{gather}
as a function of angular velocity $\Omega$ for ``frozen" $r$-modes in the $1.7\, M_\odot$ TM1C model with normal protons and neutrons. Different panels of the figure correspond to different representative stellar temperatures $T^\infty$, ranging from $10^7\,\rm K$ (panel 1) to $10^{8.8}\,\rm K$ (panel 6). In each panel, we show the roots of the residual function \eqref{Res} by red dots. For convenience, we numerate these roots sequentially through all panels, starting from the first one. Depending on the temperature $T^\infty$, we encounter that the residual function ${\rm Res}(\Omega,T^\infty)$ may have either one (panels 1, 2, and 6), two (panels 3 and 5), or three (panel 4) roots. These roots determine the boundary of the instability window at corresponding temperatures, as we demonstrate in Fig.\,\ref{windowRoots}. The figure shows the window for the same microphysical input as in Fig.\,\ref{roots}. Red dots show the points of the instability window boundary that correspond to the roots of the residual function, displayed in Fig.\,\ref{roots}. The numbering of these points is consistent with the numbering of the roots.

\newpage

\section{Conclusions}\label{Conclusion}

The results of Ref.~\cite{ofengeimetal2019} have given new impetus to the old idea that the apparent stability of rapidly rotating NSs in LMXBs can be explained by hyperonic bulk viscosity. The authors of that study reconsidered the problem by (1) using modern equations of state with $\Lambda\Xi^-$ hyperonic composition of the matter, and (2) incorporating an improved treatment of weak nonleptonic reactions that accounts for their most efficient operating channel -- through meson exchange. The results obtained in \cite{ofengeimetal2019} demonstrated that, at least for Newtonian $r$-modes, hyperonic bulk viscosity is a substantially more efficient dissipative mechanism than previously thought. Another indication that hyperonic bulk viscosity may be important for interpreting LMXB observations came from the study of relativistic $r$-modes \cite{kgk2024}, where we discovered that peculiar properties of relativistic $r$-modes in nonbarotropic matter amplify their energy dissipation through reactions.

These findings motivated us to revisit the instability windows in hyperonic stars taking relativistic effects into account. In our modeling, following \cite{ofengeimetal2019}, we employ equations of state predicting a $\Lambda\Xi^-$ hyperonic composition of the stellar mater, as opposed  to the $\Sigma^-\Lambda$ composition considered in most previous studies. Our treatment of chemical reactions in hyperonic matter also closely follows that of \cite{ofengeimetal2019} and, in particular, adopts the improved reaction rates for weak nonleptonic processes, obtained therein. Our analysis  covers the $r$-mode instability windows under various combinations of microphysical conditions. In particular, we consider the impact of weak chemical reactions on the adiabatic index $\gamma$ and estimate the possible influence of nucleon pairing. For the stellar models considered, we find that accounting for the effect of chemical reactions on the adiabatic index can shift the instability window boundary to angular velocities up to $\lesssim 10\%$ higher. Our calculations also indicate that proton superconductivity has a minor effect on the instability window. Neutron superfluidity, on the other hand, significantly suppresses $\dot{E}_\zeta$ at sufficiently low temperatures.

According to our results, relativistic effects make hyperonic bulk viscosity a much more efficient dissipation mechanism for $r$-modes than previously thought. Comparison with observations suggests that hyperonic bulk viscosity can stabilize $r$-modes even in the fastest-spinning, moderately hot neutron stars in LMXBs. Moreover, such stabilization may be possible even in the presence of nucleon pairing, provided that the critical temperatures $T_{\rm c}(r)$ for the onset of neutron superfluidity in the hyperonic core are not too high. Further, comparison of our results with their Newtonian counterparts from \cite{ofengeimetal2019} reveals that, assuming the same microphysical input, stabilization of $r$-modes in GR typically requires lower stellar masses than in Newtonian theory. For example, in the absence of pairing effects the optimal (FSU2H) Newtonian models require masses $M \gtrsim 1.7\, M_\odot$, while the corresponding relativistic models achieve stabilization at $M \gtrsim 1.6\, M_\odot$, potentially even when pairing effects are taken into account. Moreover, in relativistic (FSU2H) models, a major part of the observed sources is stabilized already at $M\gtrsim 1.5\, M_\odot$. Hence, we suggest that accounting for relativistic effects and the nonbarotropic nature of the stellar matter may be crucial for the correct interpretation of observations. 

It is important to emphasize that the instability windows we have found are quite sensitive to the assumed microphysical input (such as the equation of state, effects of superfluidity and superconductivity, etc.). This sensitivity could potentially be used to constrain the physical properties of cold, ultradense nuclear matter. Moreover, it should be noted that the available observational data (see Fig.\ \ref{representative window}) can be explained within our framework using conservative assumptions about the composition and properties of neutron star matter. Thus, stabilization of $r$-modes by nonequilibrium processes in hyperonic matter appears to be a natural mechanism for explaining the observations of rapidly rotating neutron stars in LMXBs.

The results obtained in this work rely on a number of simplifying assumptions and approximations, which warrant a brief discussion. First, we use a simplified treatment of the neutron star crust. Specifically, we neglect the effects of the crustal lattice, model the crust as a barotropic liquid, and ignore the energy dissipation within the crust as well as in the Ekman layer (i.e., the viscous boundary layer) at the crust-core interface. Such treatment is justified as long as $r$-mode frequencies are substantially higher than those of the crustal torsional modes. In this case, $r$-modes do not experience avoided crossings with crustal oscillations, and the crust, being thin compared to the core, cannot substantially influence $r$-mode eigenfunctions. Ignoring dissipation within the crust, provided by shear viscosity, is also justified, since its contribution to stabilizing $r$-modes in LMXBs is negligible compared to that of hyperonic bulk viscosity. Concerning the dissipation in the Ekman layer, it could play a significant role in stabilizing $r$-modes if the crust-core interface were sharp -- as in an idealized model of a bulk fluid rubbing against its rigid container.  In reality, however, this is likely not the case: the crust-core transition proceeds through a sequence of intermediate (pasta) phases, which smear out the boundary and suppress the formation of a well-defined viscous layer (see discussion in \cite{haskell14} and references therein). We would like to note, however, that the inclusion of the crustal lattice effects may become important in the case when $r$-mode frequencies are comparable to those of crustal torsional modes.

Second, our study ignores the modifications that nucleon pairing brings to the hydrodynamic equations. Generally, these effects require introducing additional velocity fields, describing the flows of superfluid or superconducting particles (to be more precise, superfluid/superconducting components of the considered particle species). Fortunately, hydrodynamic equations, assuming that all particle species are normal except a single superconducting one, effectively coincide with those for normal matter \cite{kgk2024}. This happens due to electromagnetic interaction, that ``attaches" the superconducting particles to the charged normal ones and, therefore, forces them to move with the same velocity. This observation suggests that our treatment of $r$-modes and their instability windows is, in fact, accurate, when neutrons are normal and protons are either normal or strongly superconducting. Neutron superfluidity, in turn, does modify hydrodynamic equations even when other particle species are normal. Therefore, our results obtained under the assumption of superfluid neutrons should be regarded as estimates, while a fully accurate treatment of the problem, of course, requires the use of superfluid hydrodynamics. Such a consideration introduces mutual friction \cite{hap2009, lm2000, ly2003, pha2009, haskell14} and resonant interaction with superfluid modes \cite{gck14a, gck2014, kg2017, dkg2019, kgd2020, kgd2021} as additional mechanisms for the damping of $r$-modes. Notably, the efficiency of mutual friction is determined by the relative velocity between the normal and superfluid components of matter, which, in turn, depends on the imbalance of chemical potentials that governs the rate of weak nonequilibrium reactions (see, e.g., \cite{lm2000}). Since the deviations from chemical equilibrium for relativistic $r$-modes are substantially larger than for Newtonian ones, we suggest that the dissipation through mutual friction might also be enhanced by relativistic effects.

Third, our calculations assume that both $\Lambda$- and $\Xi^-$-hyperons are normal. Since $\Lambda$-hyperons are involved in {\it each} process, considered in this study, their sufficiently strong superfluidity may drastically affect our conclusions, as it can suppress hyperonic bulk viscosity and, therefore, make the associated stabilization of $r$-modes in LMXBs virtually impossible. The influence of $\Xi^-$-superconductivity, in contrast, is likely weak and resembles that of proton superconductivity. Microscopically, just as proton superconductivity, $\Xi^-$-superconductivity suppresses only a limited subset of weak nonleptonic reactions [namely, processes \eqref{nXLX}-\eqref{wnpLast}], while leaving the remaining processes \eqref{wnpFirst}-\eqref{nLLL} unaffected. The associated suppression of the strong process \eqref{sp} is also unlikely to significantly influence our results (recall the discussion in the end of Sec.~\ref{SfSc effects}). Similarly to the case of protons, the motion of $\Xi^-$-hyperons is restricted by their electromagnetic interaction with other charged particle species. As a result, $\Xi^-$-superconductivity does not affect the hydrodynamic equations, if all remaining particle species are normal. We admit, however, that {\it combined} influence of $\Xi^-$ and proton superconductivity requires introducing additional velocity fields in hydrodynamic equations (see \cite{kg09} for an example of such treatment).

Fourth, we employ the Cowling approximation that neglects perturbations of the gravitational field in the hydrodynamic equations and $r$-mode energy change rates. According to \cite{lsr1990, yk1997, jc2017, kgk2024, kgk2022_1, kgk2022_2}, this approximation is believed to provide reasonable estimates of the $r$-mode properties. Moreover, our preliminary results indicate that, in the $\Omega \to 0$ limit, $r$-modes in full GR are governed by {\it exactly} the same equations as those of the hydrodynamic approach, taken in the appropriate limit. This finding confirms that the Cowling approximation accurately captures relativistic $r$-mode behavior at slow rotation rates.

Next, we neglect the oblateness of the star induced by rotation [i.e., the $O(\Omega^2/\Omega_{\rm K}^2)$ terms in \eqref{geometry}–\eqref{redshift}]. We adopt this approximation primarily to simplify the equations and admit that it should be discarded in the most accurate consideration of the problem. Nevertheless, we expect this approximation to have only a minor impact on the $r$-mode properties and their instability windows. Indeed, we verified that accounting for oblateness does not qualitatively affect the mathematical structure of the $r$-mode equations both in GR and in Newtonian theory. We have also verified that it does not modify the expressions \eqref{All Edot} for the relativistic $r$-mode energy change rates and their Newtonian counterparts. Additionally, we note that, in the case of Newtonian $r$-modes, oblateness mostly influences their eigenfrequency correction $\sigma^{(1)}$, while leaving the toroidal function $T_m(r)$ essentially unchanged. At the same time, for Newtonian $r$-modes, oblateness {\it does} alter the  derivative $h'(r)$ showing the effect of matter nonbarotropicity, but in such a way that its absolute value $|h'(r)|$, that determines dissipation through bulk viscosity, remains of approximately the same order of magnitude as in a spherical model. Also note that, since accounting for oblateness preserves the mathematical structure of the equations, it clearly preserves the peculiar properties of relativistic $r$-modes. These observations indicate that accounting for oblateness should not substantially affect our results.

Finally, our calculations formally treat the framedragging effect $\omega(r)$ [see the metric tensor \eqref{geometry}], responsible for peculiar behavior of relativistic $r$-modes, as weak. This approximation is somewhat forced, as, to the best of our knowledge, it currently remains perhaps the most viable approach for modeling relativistic $r$-modes in nonbarotropic matter without encountering the continuous spectrum problem (see our literature review in \cite{kgk2022_1})%
\footnote{
The only alternative approach to calculating $r$-mode eigenfunctions we are aware of is the straightforward numerical consideration of Yoshida \& Lee \cite{yl2002}, which is rather cumbersome, as it does not use the slow-rotation approximation.
}. 
We expect this approximation to accurately describe $r$-mode behavior at sufficiently slow rotation rates, when $r$-modes are localized in the outer layers of the star where the framedragging effect, indeed, can be treated as weak. In the deeper layers of the star, especially those closer to its center, the function $\omega(r)$ can no longer be considered small, and the approximation becomes less accurate. Nevertheless, it still provides a reasonable estimate of the key relativistic $r$-mode properties -- namely, their nonanalytic behavior and characteristic scaling $\xi^r_{m+1}(r)\sim \Omega\, T_m(r)$. In fact, we are currently working on extending the theory to the case of a realistic (i.e., not necessarily weak) framedragging effect. Our preliminary results indicate that relaxing this approximation preserves these key features of relativistic $r$-modes. We will provide all the relevant details in our forthcoming publication, which is expected in the near future.


\section*{Acknowledgements}
The authors are grateful to A.I.~Chugunov for stimulating discussions and critical comments. This research was started during a long-term visit by the authors to the Weizmann Institute of Science (WIS). We acknowledge the support of the visit by the Simons Foundation and WIS. The authors are grateful to the Department of Particle Physics \& Astrophysics at WIS for their hospitality and excellent working conditions. The authors also acknowledge the support of this work by Russian Science Foundation [Grant 22-12-00048-P].

\appendix


\section{Modeling relativistic $r$-modes}\label{Modeling r-modes}

Since the $l=m=2$ $r$-mode is expected to be the most CFS-unstable, in what follows, we restrict ourselves to the $l=m$ case. We remind the reader that we treat the NS crust as barotropic and NS core as nonbarotropic. Nonbarotropicity of the matter significantly modifies the mathematical properties of the oscillation equations, motivating us to consider the crust and core equations separately. Below we briefly summarize the results from our original papers \cite{kgk2022_1, kgk2022_2}, devoted to hydrodynamic modeling of relativistic $r$-modes.

By considering nondissipative hydrodynamic equations \eqref{nondissipative equations}, one can show that in the crust the $r$-mode eigenfunctions are determined from the following system:
\begin{gather}
\left\{
\begin{gathered}
T_m^{\rm crust}(r)=\const \cdot r^m e^{-(m-1)[\nu(r)-\nu(R)]} \hfill \\
\frac{d}{dr}\xi^{\rm crust}_{m+1}(r)+g_1(r)\xi_{m+1}^{\rm crust}(r)+ \hfill \\
\hfill + \biggl\{g_{21}(r)\biggl[\sigma^{(1)}+\frac{2\omega(r)}{m+1}\biggr]+\Omega^2 g_{22}(r)\biggr\}T^{\rm crust}_m(r)=0, \label{system crust} \raisetag{1.5cm}
\end{gathered}
\right.
\end{gather}
with supplementary coefficients $g_1(r)$, $g_{21}(r)$, $g_{22}(r)$ defined by
\begin{gather}
\begin{gathered}
g_1(r)=\frac{m+3}{r}-g(r)\left[m+1+\frac{c^2}{c_s^2(r)}\right]+\lambda'(r), \hfill \\
g_{21}(r)=-\frac{(m+1)^2(2m+3)}{2r(2m+1)}, \hfill \\
g_{22}(r)=\frac{r\, m \left[(m+1)(3m^2+2m-9) c_s^2(r)-8 c^2\right]e^{-2\nu(r)}}{2 c^2 (m+1)^2 (2 m+1) c_s^2(r)}. \hfill
\end{gathered}
\raisetag{1.8cm}
\end{gather}
Here we introduced the (divided by $c^2$) gravitational acceleration $g(r)$ and the crustal equilibrium speed of sound $c_{s}(r)$ as
\begin{gather}
\begin{gathered}
g(r)=\nu'(r)=-\frac{p_0'(r)}{w_0(r)}, \hfill
\\
c_{s}(r)=c\sqrt{\biggl(\frac{dp}{d\varepsilon}\biggr)_0}=c\sqrt{\frac{p_0'(r)}{\varepsilon_0'(r)}}. \hfill
\end{gathered}
\end{gather}

Next, one can show that in the nonbarotropic core hydrodynamic equations \eqref{nondissipative equations} reduce to the system of ordinary differential equations that can be written schematically in the following form:
\begin{gather}
 \label{system core} 
\left\{
\begin{gathered}
\biggl[C_{1}(r)\frac{d}{d r}+C_{2}(r)\biggr]\xi^{\rm core}_{m+1}(r)+ \hfill \\
\hfill +\biggl[\Omega^2 C_{3}(r)+\sigma^{(1)}+\frac{2\omega(r)}{m+1}\biggr]T^{\rm core}_m(r)=0 \\
\biggl[\frac{d}{dr}+G_1(r)\biggr]T^{\rm core}_m(r)+\frac{G_2(r)}{\Omega^2}\xi^{\rm core}_{m+1}(r)=0.
\end{gathered}
\right.
\end{gather}
The coefficients $C_{1,2,3}(r)$ and $G_{1,2}(r)$ in this system are given by the following expressions:
\begin{gather}
\begin{gathered}
C_1(r)=-\frac{2rk^+_{m+1}}{(m+1)^2}, \hfill \\
C_2(r)=\frac{k^{+}_{m+1}}{(m+1)^2}\biggl[2rg(r)(m+1)-F(r)-2m-1\biggr],  \hfill  \\
C_3(r)=\frac{r^2 e^{-2\nu(r)}}{c^2 m^2(m+1)^2}\biggl[\gamma_2+8 m\gamma_1 \frac{w_0(r)}{\gamma(r) p_0(r)}\biggr], \hfill  \\
G_1(r)=A(r)+g(r)(m-1)-\frac{m}{r},  \hfill  \\
G_2(r)=\frac{A(r)g(r)c^2(m+1)^2 e^{2\nu(r)}}{4 \ r \ m \ k^-_m},  \hfill 
\end{gathered}
\end{gather}
with
\begin{gather}
\begin{gathered}
k^+_L=\frac{L+m}{2L+1}, \qquad k^-_L=\frac{L-m+1}{2L+1}, \hfill \\
\gamma_1=\frac{m^2}{(m+1)^2(2m+3)}, \hfill \\
\gamma_2=\frac{m^3[9-m(3m+2)]}{(m+1)(2m+3)}, \hfill \\
F(r)=2 r \biggl[\lambda'(r)-\frac{g(r)w_0(r)}{\gamma(r)p_0(r)}\biggr]+5. \hfill
\end{gathered}
\end{gather}
For brevity, in these formulas, we omit the dependence of the adiabatic index $\gamma(r,\Omega,T^\infty)$, Schwarzschild discriminant $A(r,\Omega,T^\infty)$, and the relevant supplementary functions and coefficients on $\Omega$ and $T^\infty$.

Crust \eqref{system crust} and core \eqref{system core} equations cannot be considered independently, as they are coupled through the shared eigenfrequency correction $\sigma^{(1)}$ and boundary conditions.  Specifically, we require the continuity of the energy and momentum currents at the crust-core interface $r=r_{\rm cc}$, which leads to
\begin{gather}
T^{\rm core}_{m}(r_{\rm cc})=T^{\rm crust}_{m}(r_{\rm cc}), \quad \xi^{\rm core}_{m+1}(r_{\rm cc})=\xi^{\rm crust}_{m+1}(r_{\rm cc}).
\end{gather}
We also require the regularity (i.e., nondivergent behavior) of the $r$-mode eigenfunctions at the stellar center $r=0$, and vanishing total pressure at the perturbed NS surface. The latter condition is equivalent to 
\begin{gather}
\xi^{\rm crust}_{m+1}(R)=-\frac{4\Omega^2 m R e^{-2\nu(R)}}{c^2 (m+1)^2 (2m+1)g(R)}T^{\rm crust}_m(R).
\end{gather}
All the listed boundary conditions can be satisfied {\it simultaneously} only for the specific values of the eigenfrequency correction $\sigma^{(1)}$, corresponding to the global $r$-mode solutions. These solutions are suitably distinguished by the number of nodes (zeros) of the toroidal function $T_m(r)$ inside the star. In this paper, we focus on the nodeless $r$-mode, which is often referred to as {\it fundamental}.

\newpage



%

\end{document}